\documentclass[aps,nofootinbib,floatfix,showpacs,preprintnumbers,twocolumn,superscriptaddress]{revtex4} 
\usepackage{graphicx}
\usepackage{bm}
\usepackage{mathrsfs}
\usepackage{float}
\usepackage{tabularx} 
\usepackage{amsmath}
\usepackage{epstopdf}
\usepackage{xcolor}
\usepackage{hyperref}
\usepackage{natbib}
\input epsf

\def\lsim{\mathrel{\raise.3ex\hbox{$<$\kern-.75em\lower1ex\hbox{$\sim$}}}}
\def\gsim{\mathrel{\raise.3ex\hbox{$>$\kern-.75em\lower1ex\hbox{$\sim$}}}}

\def\cmm2{{\,\rm cm^{-2}}}
\def\cm2{{\,{\rm cm}^2}}
\def\cmm3{{\,{\rm cm}^{-3}}}
\def\gcmm3{{\,{\rm g\,cm^{-3}}}}

\def\fun#1#2{\lower3.6pt\vbox{\baselineskip0pt\lineskip.9pt
  \ialign{$\mathsurround=0pt#1\hfil##\hfil$\crcr#2\crcr\sim\crcr}}}

\def\be{\begin{equation}}
\def\ee{\end{equation}}
\def\bea{\begin{eqnarray}}
\def\eea{\end{eqnarray}}

\begin{document}

\title{
Searching for low-mass dark matter particles with a massive Ge bolometer operated above-ground}

\author{E. Armengaud}
\affiliation{IRFU, CEA, Universit\'{e} Paris-Saclay, F-91191 Gif-sur-Yvette, France}
\author{C. Augier}
\affiliation{Univ Lyon, Universit\'e Lyon 1, CNRS/IN2P3, IPN-Lyon, F-69622, Villeurbanne, France}
\author{A.~Beno\^{i}t}
\affiliation{Institut N\'{e}el, CNRS/UJF, 25 rue des Martyrs, BP 166, 38042 Grenoble, France}
\author{A.~Benoit}
\affiliation{Univ Lyon, Universit\'e Lyon 1, CNRS/IN2P3, IPN-Lyon, F-69622, Villeurbanne, France}
\author{L.~Berg\'{e}}
\affiliation{CSNSM, Univ. Paris-Sud, CNRS/IN2P3, Universit\'{e} Paris-Saclay, 91405 Orsay, France}
\author{J.~Billard}\email{j.billard@ipnl.in2p3.fr}
\affiliation{Univ Lyon, Universit\'e Lyon 1, CNRS/IN2P3, IPN-Lyon, F-69622, Villeurbanne, France}
\author{A.~Broniatowski}
\affiliation{CSNSM, Univ. Paris-Sud, CNRS/IN2P3, Universit\'{e} Paris-Saclay, 91405 Orsay, France}
\author{P.~Camus}
\affiliation{Institut N\'{e}el, CNRS/UJF, 25 rue des Martyrs, BP 166, 38042 Grenoble, France}
\author{A.~Cazes}
\affiliation{Univ Lyon, Universit\'e Lyon 1, CNRS/IN2P3, IPN-Lyon, F-69622, Villeurbanne, France}
\author{M.~Chapellier}
\affiliation{CSNSM, Univ. Paris-Sud, CNRS/IN2P3, Universit\'{e} Paris-Saclay, 91405 Orsay, France}
\author{F.~Charlieux}
\affiliation{Univ Lyon, Universit\'e Lyon 1, CNRS/IN2P3, IPN-Lyon, F-69622, Villeurbanne, France}
\author{D.~Ducimeti\`ere}
\affiliation{Univ Lyon, Universit\'e Lyon 1, CNRS/IN2P3, IPN-Lyon, F-69622, Villeurbanne, France}
\author{L.~Dumoulin}
\affiliation{CSNSM, Univ. Paris-Sud, CNRS/IN2P3, Universit\'{e} Paris-Saclay, 91405 Orsay, France}
\author{K.~Eitel}
\affiliation{Karlsruher Institut f\"{u}r Technologie, Institut f\"{u}r Kernphysik, Postfach 3640, 76021 Karlsruhe, Germany}
\author{D.~Filosofov}
\affiliation{JINR, Laboratory of Nuclear Problems, Joliot-Curie 6, 141980 Dubna, Moscow Region, Russian Federation}
\author{J.~Gascon}
\affiliation{Univ Lyon, Universit\'e Lyon 1, CNRS/IN2P3, IPN-Lyon, F-69622, Villeurbanne, France}
\author{A.~Giuliani}
\affiliation{CSNSM, Univ. Paris-Sud, CNRS/IN2P3, Universit\'{e} Paris-Saclay, 91405 Orsay, France}
\author{M.~Gros}
\affiliation{IRFU, CEA, Universit\'{e} Paris-Saclay, F-91191 Gif-sur-Yvette, France}
\author{M. De~J\'{e}sus}
\affiliation{Univ Lyon, Universit\'e Lyon 1, CNRS/IN2P3, IPN-Lyon, F-69622, Villeurbanne, France}
\author{Y.~Jin}
\affiliation{Laboratoire de Photonique et de Nanostructures, CNRS, Route de Nozay, 91460 Marcoussis, France}
\author{A.~Juillard}
\affiliation{Univ Lyon, Universit\'e Lyon 1, CNRS/IN2P3, IPN-Lyon, F-69622, Villeurbanne, France}
\author{M.~Kleifges}
\affiliation{Karlsruher Institut f\"{u}r Technologie, Institut f\"{u}r Prozessdatenverarbeitung und Elektronik, Postfach 3640, 76021 Karlsruhe, Germany}
\author{R.~Maisonobe}
\affiliation{Univ Lyon, Universit\'e Lyon 1, CNRS/IN2P3, IPN-Lyon, F-69622, Villeurbanne, France}
\author{S.~Marnieros}
\affiliation{CSNSM, Univ. Paris-Sud, CNRS/IN2P3, Universit\'{e} Paris-Saclay, 91405 Orsay, France}
\author{D.~Misiak}
\affiliation{Univ Lyon, Universit\'e Lyon 1, CNRS/IN2P3, IPN-Lyon, F-69622, Villeurbanne, France}
\author{X.-F.~Navick}
\affiliation{IRFU, CEA, Universit\'{e} Paris-Saclay, F-91191 Gif-sur-Yvette, France}
\author{C.~Nones}
\affiliation{IRFU, CEA, Universit\'{e} Paris-Saclay, F-91191 Gif-sur-Yvette, France}
\author{E.~Olivieri}
\affiliation{CSNSM, Univ. Paris-Sud, CNRS/IN2P3, Universit\'{e} Paris-Saclay, 91405 Orsay, France}
\author{C.~Oriol}
\affiliation{CSNSM, Univ. Paris-Sud, CNRS/IN2P3, Universit\'{e} Paris-Saclay, 91405 Orsay, France}
\author{P.~Pari}
\affiliation{IRAMIS, CEA, Universit\'{e} Paris-Saclay, F-91191 Gif-sur-Yvette, France}
\author{B.~Paul}
\affiliation{IRFU, CEA, Universit\'{e} Paris-Saclay, F-91191 Gif-sur-Yvette, France}
\author{D.~Poda}
\affiliation{CSNSM, Univ. Paris-Sud, CNRS/IN2P3, Universit\'{e} Paris-Saclay, 91405 Orsay, France}
\author{E.~Queguiner}
\affiliation{Univ Lyon, Universit\'e Lyon 1, CNRS/IN2P3, IPN-Lyon, F-69622, Villeurbanne, France}
\author{S.~Rozov}
\affiliation{JINR, Laboratory of Nuclear Problems, Joliot-Curie 6, 141980 Dubna, Moscow Region, Russian Federation}
\author{V.~Sanglard}
\affiliation{Univ Lyon, Universit\'e Lyon 1, CNRS/IN2P3, IPN-Lyon, F-69622, Villeurbanne, France}
\author{B.~Siebenborn}
\affiliation{Karlsruher Institut f\"{u}r Technologie, Institut f\"{u}r Kernphysik, Postfach 3640, 76021 Karlsruhe, Germany}
\author{L.~Vagneron}
\affiliation{Univ Lyon, Universit\'e Lyon 1, CNRS/IN2P3, IPN-Lyon, F-69622, Villeurbanne, France}
\author{M.~Weber}
\affiliation{Karlsruher Institut f\"{u}r Technologie, Institut f\"{u}r Prozessdatenverarbeitung und Elektronik, Postfach 3640, 76021 Karlsruhe, Germany}
\author{E.~Yakushev}
\affiliation{JINR, Laboratory of Nuclear Problems, Joliot-Curie 6, 141980 Dubna, Moscow Region, Russian Federation}
\author{A.~Zolotarova}\altaffiliation[Present address: ] {CSNSM, Univ. Paris-Sud, CNRS/IN2P3, Universit\'{e} Paris-Saclay, 91405 Orsay, France}
\affiliation{IRFU, CEA, Universit\'{e} Paris-Saclay, F-91191 Gif-sur-Yvette, France}

\author{(EDELWEISS Collaboration)} 
\noaffiliation
\author{B. J. Kavanagh}\email{b.j.kavanagh@uva.nl}
\affiliation{GRAPPA, University of Amsterdam, Science Park 904, 1098 XH Amsterdam, The Netherlands}

\date{\today}
\smallskip
\begin{abstract}
The EDELWEISS collaboration has performed a search for dark matter particles with masses below the GeV-scale with a 33.4-g germanium cryogenic detector operated in a surface lab. The energy deposits were measured using a neutron-transmutation-doped Ge thermal sensor with a 17.7~eV (RMS) baseline heat energy resolution leading to a 60~eV analysis energy threshold. Despite a moderate lead shielding and the high-background environment, the first sub-GeV spin-independent dark matter limit based on a germanium target has been achieved. The experiment provides the most stringent, nuclear recoil based, above-ground limit on spin-independent interactions above 600~MeV/c$^{2}$. The experiment also provides the most stringent limits on spin-dependent interactions with protons and neutrons below 1.3~GeV/c$^{2}$. Furthermore, the dark matter search results were studied in the context of Strongly Interacting Massive Particles, taking into account Earth-shielding effects, for which new regions of the available parameter space have been excluded. Finally, the dark matter search has also been extended to interactions via the Migdal effect, resulting for the first time in the exclusion of particles with masses between 45 and 150~MeV/c$^{2}$ with spin-independent cross sections ranging from $10^{-29}$ to $10^{-26}$~cm$^2$.
\end{abstract}
\maketitle

\section{Introduction}
\label{sec:intro}

Various cosmological observations indicate that $26 \%$ of the energy density of the Universe is in the form of cold, non-baryonic, dark matter (DM)~\cite{planck2018-vi}. Weakly Interacting Massive Particles (WIMPs) are suitable cold DM candidates; they arise in extensions of the Standard Model of Particle Physics, such as Supersymmetry, and are naturally produced in the early Universe with the correct abundance (for reviews see e.g.~Refs.~\cite{Jungman,Bertone}). WIMPs from the Milky Way's dark matter halo can be detected directly on Earth, via the keV-scale recoils produced when they elastically scatter off nuclei~\cite{Goodman:1984dc,annual}.
In recent decades, significant advances have been made in the search for WIMPs in the GeV/c$^{2}$- to TeV/c$^{2}$-range that is natural for Supersymmetry~\cite{xenon1t,lux,pandax}. However, in the light of the absence of signal in that region there is an increasing interest in DM particles in the GeV/c$^{2}$ and sub-GeV/c$^{2}$ mass range~\cite{Essig,Cheung,Hooper,Falkowski,Petraki,Zurek,Bertone:2018krk}. 
These searches require experimental thresholds as low as a few tens of eV, a performance that can be attained by cryogenic detectors~\cite{cresst,cdmslite}.
A particular advantage of such detector technology is that the thermal signal is not affected by the strong quenching effects that tend to severely reduce the amplitude of ionization or scintillation signals at low energy.

This paper describes the results obtained by the EDELWEISS collaboration with a 33.4-g Ge detector demonstrating that such a device equipped with a neutron-transmutation-doped Ge (Ge-NTD) sensor~\cite{NTD} can reach the sensitivity to probe the sub-GeV domain. 
As a proof of the relevance of this technology, it is used in a dedicated EDELWEIS-Surf run devoted to an above-ground search for DM particles.
Such a search is bound to be limited by the large background induced by cosmic-ray interactions, but has the advantage of being able to probe models beyond the simple WIMP paradigm by considering relatively large values for the DM-nucleon cross sections.
For sufficiently large values, each DM particle will typically interact many times in the atmosphere, Earth and shielding before reaching an underground detector. 
A DM particle loses energy with each interaction and may therefore arrive at the detector with insufficient energy to be observed above threshold \cite{Starkman:1990nj,Collar1992,Collar1993,Hasenbalg:1997hs,Kouvaris:2014lpa,Kouvaris:2015laa,Bernabei:2015nia,Kavanagh:2016pyr}. Above-ground searches minimize this `stopping' effect in the Earth and therefore provide good sensitivity to large DM-nucleon cross sections. While most current constraints on strongly interacting massive particles (SIMPs\footnote{We use the term `SIMP' here to refer to DM with large scattering cross sections with ordinary matter. However, we note that the term SIMP may also be used for hidden sector DM with strong self-interactions \cite{Hochberg:2014dra,Hochberg:2014kqa}.}) rely on a reanalysis of public data \cite{Albuquerque:2003ei,Davis:2017noy,Kavanagh:2017cru}, we perform a dedicated search that fully takes into account the detailed detector response, presenting both the smallest and largest cross sections to which the experiment is sensitive.

We also present a search for WIMP-induced nuclear recoils which are accompanied by the ionization of an atomic electron~\cite{Vergados:2004bm,Moustakidis:2005gx,Bernabei:2007jz}. 
The total energy in the nuclear recoil and ionization is typically larger than what can be deposited by elastic nuclear recoils at a given DM mass \cite{Ibe:2017yqa}. 
This `Migdal effect' therefore allows one to probe lighter DM particles (for a fixed energy threshold) at the expense of a smaller event rate. 
Constraints using the Migdal effect have been reported previously for NaI crystal detectors~\cite{Bernabei:2007jz} and liquid Xenon detectors~\cite{Dolan:2017xbu,Akerib:2018hck}. 
Here, we present the first Migdal limit using a Germanium target.

The paper is organized as follows. In Sec.~\ref{sec:surf}, we outline the experimental setup. In Sec.~\ref{sec:DMsearch}, we give details of the dark matter search, including data processing, detector calibration and data analysis. In Sec.~\ref{sec:results}, we present limits on weakly- and strongly-interacting DM, for both elastic nuclear recoils and the inelastic Migdal effect. The resulting exclusion regions are summarized in Fig.~\ref{fig:Limit}. Finally, we conclude in Sec.~\ref{sec:conclusion}.

\section{EDELWEISS-Surf}
\begin{figure*}[!t]
\begin{center}
\includegraphics[width=0.49\textwidth,angle=0]{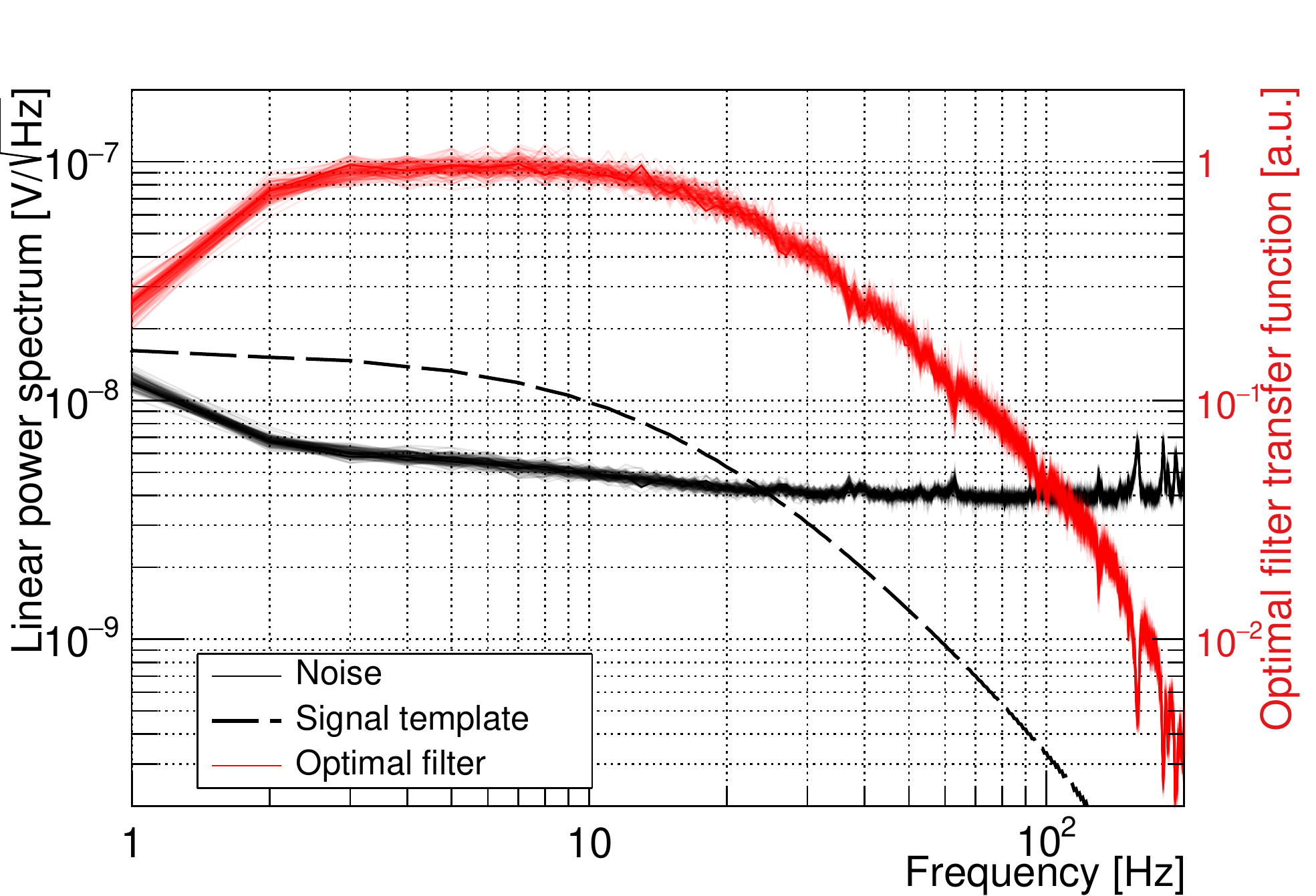}
\includegraphics[width=0.49\textwidth,angle=0]{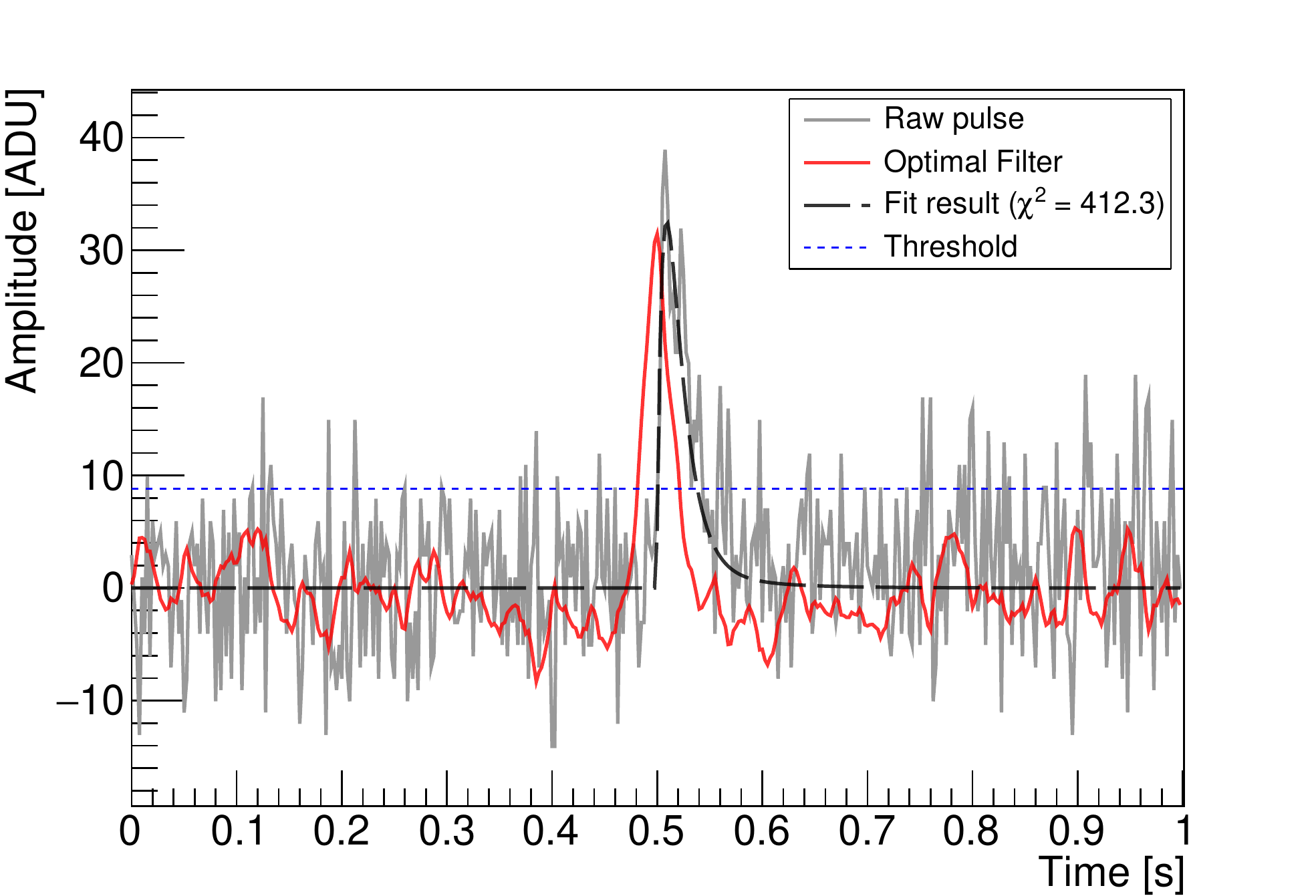}
\caption{{\bf Left}: Hourly-averaged noise Power Spectral Densities (PSD) (black curves), detector signal bandwidth (black dashed line), and resulting optimal filter transfer functions (red curves) as a function of frequency, for the six days of data acquisition. The 137 separate PSDs and transfer functions are overlayed. {\bf Right}: Example of a 200 eV pulse: unfiltered raw trace (grey solid line) and output of the optimal filter (red solid line).
The trigger level at 3$\sigma$ is shown as the blue dotted line.
 The result of the pulse fitting procedure, with a $\chi^2$/ndf = 1.03, is shown as the black long-dashed line.} 
\label{fig:Pulse}
\end{center}
\end{figure*} 
\label{sec:surf}
\subsection{Detector}
\label{sec:detector}
The detector prototype consists of a cylindrical high-purity Ge crystal of 20 mm diameter and 20 mm height, corresponding to a total mass of 33.4-g. 
The thermal sensor design has been optimized for enhanced heat energy resolution. It consists of a Ge-NTD of $2\times 2\times 0.5$ mm$^3$, glued on the top surface of the crystal, weakly thermally coupled to the copper housing of the detector thanks to gold wire bonds connecting its electrodes to two gold pads on a Kapton tape.   
With a total Ge-NTD electrode surface of 2 mm$^2$ this weak thermal link is about 2.1~nW/K which is sub-dominant with respect to the electron-phonon coupling of 6.7~nW/K ensuring that the detector properly integrates all of the heat signal~\cite{pyle}.
The crystal is held by six PTFE clamps (three on each side) in order to ensure the mechanical constraints on all three axes of displacement and to minimize the stress due to PTFE elasticity at low temperatures. Unlike the usual EDELWEISS-III FID800 detectors~\cite{edwtech}, this detector prototype has only one heat channel and no ionization readout. Therefore, a discrimination between nuclear and electron recoils is not possible. However, as there is no electric field applied across the crystal, the detector acts as a true calorimeter measuring the deposited energy of the recoiling particle independently of its type (nuclear or electronic recoil).
Quenching effects on the heat energy scale for nuclear recoils in Ge cryogenic detectors have been shown to be very small~\cite{quenching-edw,quenching-cdms}, and are therefore neglected hereafter. 

\subsection{Experimental setup}

The dark matter search has been performed in the dry dilution cryostat of the \textit{Institut de Physique Nucl\'eaire de Lyon} installed in a surface building with negligible overburden, see Sec.~\ref{sec:SIMP}. The cryostat is a Hexadry-200 commercially available from Cryoconcept~\cite{cryoconcept}, 
which has been upgraded to reduce the vibration levels of the mixing chamber by mechanically decoupling the cold head of the pulse tube cryocooler from the dilution unit~\cite{cryovib}. 
The vibrations at the detector level were further mitigated with the use of a dedicated suspended tower~\cite{suspendedtower}. The latter consists in a 25 cm long elastic pendulum, attached to the 1 K stage by a Kevlar string and a stainless steel spring with an elastic constant of 240 N/m, holding the detector tower situated below the mixing chamber at 10 mK. The detector tower is thermally anchored to an intermediate holding structure, via supple copper braids, which also hosts the connectors for the detector readout. This suspended tower design reduces detector vibrations at the sub-$\mu$g/$\sqrt{\text{Hz}}$ level, with displacements in the order of a few nanometers (RMS) in all three axes, leading to substantial gains in energy resolutions as demonstrated in Ref.~\cite{suspendedtower}. The cold and warm electronics are those described in Ref.~\cite{edwtech}, with a first Bi-FET preamplifier stage at 100 K and a second stage amplifier at 300 K.

The cryostat is surrounded by a 10 cm thick cylindrical lead shield covering a solid angle of $\sim$70\% of $4\pi$ around the detector. The materials used for the cryostat construction were not selected for low radioactivity, with the exception of the replacement of the standard glass fiber rods used by Cryoconcept by stainless steel ones, shown to have much less radioactive contamination. 

Finally, an $^{55}$Fe calibration source was glued on the inner part of the detector's copper housing and facing the crystal surface opposite to the side on which is glued the Ge-NTD.

\section{Dark Matter search}
\label{sec:DMsearch}

The dark matter search started two weeks after the mixing chamber reached its base temperature of 10 mK. 
During these first two weeks, the thermal response of the detector was studied, and its working point was optimized. 
The best heat energy resolution achieved was $~17.7$~eV (RMS), with the temperature of the suspended tower regulated at 17~mK and the Ge-NTD biased at 1~nA, 
leading to a steady state resistance of 3.4 M$\Omega$. 
After the optimization period, it was decided to record six days of data in these conditions from May 22$^{\mathrm{nd}}$ until May 27$^{\mathrm{th}}$  2018. It was decided beforehand to blind a 24-hour long data period started at 5 pm on May 26$^{\mathrm{th}}$. The remaining 5-days worth of data were then used to both tune the analysis procedure and selection cuts, and to build a data driven background model, in order to derive dark matter constraints from a blind analysis.

\subsection{Data processing}

The data acquisition used here is the same as from the EDELWEISS-III experiment located at the {\it Laboratoire Souterrain de Modane}~\cite{edwtech}.
In order to cancel common electronic noise sources and reduce microphonics, the voltage drop across the Ge-NTD is measured differentially and the current across it is modulated from positive to negative values, following a square wave function. A modulation frequency $f_s$ of 400~Hz was chosen as it resulted in the best achievable signal-to-noise ratio in the experimental conditions considered here. 
The data are recorded continuously at this effective sampling frequency $f_s$, in so-called {\it stream mode}, such that there is no online trigger, unlike the standard EDELWEISS-III data acquisition. Instead, pulse signals are identified offline thanks to a dedicated signal processing pipeline, described in the following, which optimally filters the data based on the frequency dependence of the observed signal and noise. In order to avoid any noise structures at frequencies below the analysis range, from 1~Hz to 200~Hz, a second-order Butterworth numerical filter with a cut-off frequency of 2~Hz has been applied to the entire data stream before any selection of noise traces and triggers. Unless otherwise stated, the remaining part of the data processing is based on this pre-filtered data stream.

\subsubsection{Noise PSD estimation}
\begin{figure*}[!t]
\begin{center}
\includegraphics[width=0.49\textwidth,angle=0]{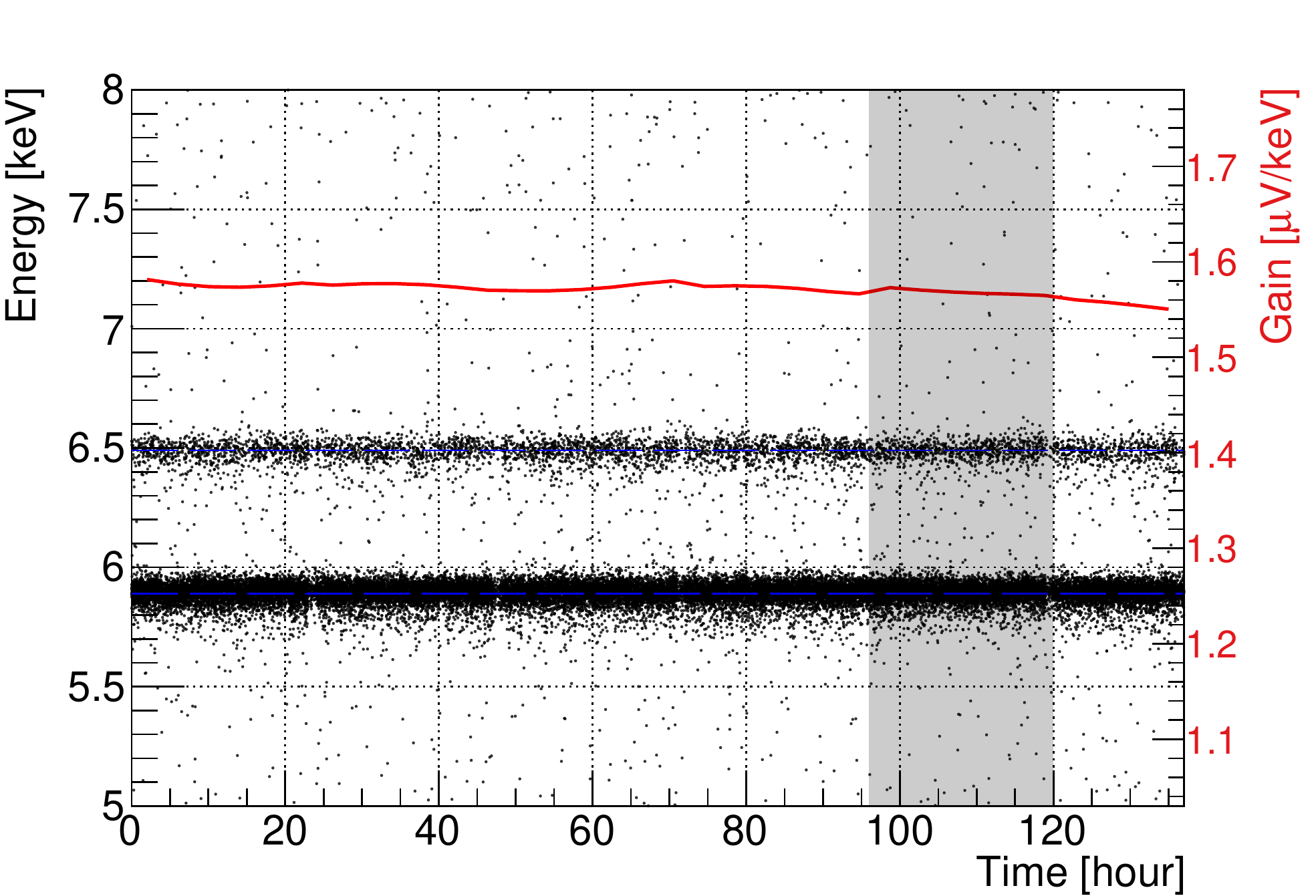}
\includegraphics[width=0.49\textwidth,angle=0]{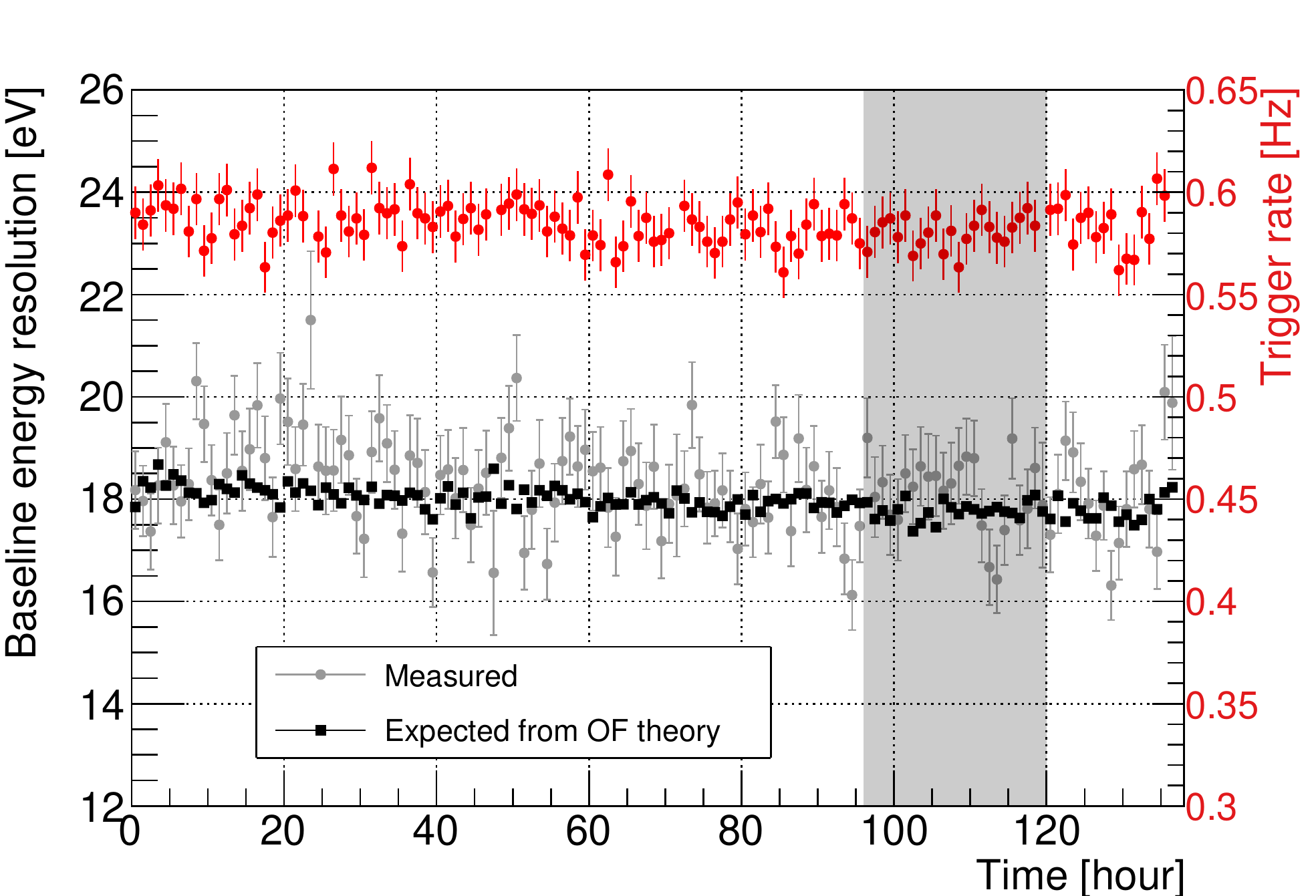}
\caption{{\bf Left}: Event energy distribution between 5 and 8 keV as a function of time. The horizontal bands at 5.90 and 6.49 keV correspond to the $K_\alpha$ and $K_\beta$ X-ray lines, respectively, of Mn emitted by the $^{55}$Fe source. The data have been corrected for the measured time-evolution of the detector gain as a function of time, shown as the red line, and corresponding to the right-hand axis. 
{\bf Right}: Baseline heat energy resolution (RMS) in eV as a function of time. The grey dots are the values derived from a fit to the energy distributions of the noise event selection, and the black squares are those derived from the ratio of the signal and noise PSDs. The corresponding trigger rates in Hz are shown as red dots. Each data point corresponds to one hour. The grey shaded region in both panels corresponds to the interval of the blinded data set.
 } 
\label{fig:Calibration}
\end{center}
\end{figure*} 
The noise Power Spectral Density (PSD) of the heat channel for an hour of recorded data is recursively determined from a random sample of 1-s time traces uniformly selected throughout the entire data stream for that hour. After an initial selection, based on their RMS dispersion, the removal of time traces containing pulses proceeds recursively by computing their individual frequency-based $\chi^2$ with respect to the averaged PSD. The procedure stops once the mean $\chi^2$, from both the pre-filtered and raw data stream, from all selected time traces is compatible with the expected value of $N_s$ = 400, corresponding to the number of time samples per trace. An average noise PSD is thus determined for each of the 137 hours that comprises the entire data set.
These PSDs, corrected for the 2~Hz filter gain, are overlaid on Fig.~\ref{fig:Pulse} (left panel).
The small dispersion shows that, despite its surface operation, the noise is very stable over the entire 6-days long run.

The PSD plateaus at a value of 4 nV/$\sqrt{\text{Hz}}$ above 20 Hz,
with very little electromagnetic pickup contributions. This value is very well explained by the quadratic addition of the Johnson noise of the Ge-NTD (1.8 nV/$\sqrt{\text{Hz}}$) and the current noise of the Bi-FET preamplifier. 
The slow rise in the noise level below 20 Hz, reaching a value of 6 nV/$\sqrt{\text{Hz}}$ at 2~Hz, is due to internal thermal fluctuation noise from the detector~\cite{thermModel}.

\subsubsection{Offline trigger} 
\label{sec:trigger}

The offline trigger is based on a match filtered approach~\cite{trigger} designed to maximize the signal-to-noise ratio in the estimation of the signal pulse amplitude. 
The filter $H(f_i)$ considered hereafter is derived from the measured frequency dependence of the signal-to-noise ratio
and is therefore defined as: 
\begin{equation}
H(f_i) = h\frac{s^{*}(f_i)}{J(f_i)}e^{-j2\pi f_i t_M}
\label{eq:filter},
\end{equation}
where $s^{*}(f_i)$ is the complex conjugate of the signal template (shown as the black dashed line in the left panel of Fig.~\ref{fig:Pulse}), $J(f_i)$ is the noise PSD in V$^2$/Hz (black solid lines), $t_M$ corresponds to the time position of the pulse template maximum, and $h$ is a normalization constant that preserves the amplitude of the pulse signal such that
\begin{equation}
h = \left(\sum_i \frac{|s(f_i)|^2}{J(f_i)}\right)^{-1},
\end{equation}
where $f_i$ varies between -$f_s/2$ to +$f_s$/2.
An optimally matching filter is determined from each of the recorded 137 hours. 
Their moduli $|H(f_i)|$ are also shown as the red solid lines in the left panel of Fig.~\ref{fig:Pulse}. 
As one can derive from the latter, only the lowest frequencies (below ~50 Hz) are relevant to recover the observed pulse amplitude. 
The filter from Eq.~\ref{eq:filter} is applied to the data using the numerical procedure described in Ref.~\cite{trigger}. 
As an example, the right panel of Fig.~\ref{fig:Pulse} shows a 200 eV event prior (grey solid line) and post (red solid line) filtering.
Also shown is the best pulse fitting solution following the event processing procedure described in Sec.~\ref{sec:processing}. Candidate events are identified when the filtered data exceed a given threshold level which has been defined in terms of a fixed number $n$ of the baseline energy resolution $\sigma_{OF}$, where $\sigma_{OF}=\sqrt{h}$.
The value of $n$ was chosen such that the rate of noise induced triggers is significantly smaller than the rate of physical events, which is about $\sim$1.3~Hz. The dependence on the value of $n$ of the rate of noise induced triggers has been evaluated by simulating a 24-hour long stream of fake data using the observed noise PSD, without injecting any signal pulses, and applying the same triggering procedure as for real data. We chose $n=3$ as it resulted in an optimum between a low energy threshold and a reasonably low expected noise induced trigger rate of 0.15 Hz. With the observed value of $\sigma_{OF}$ of 18 eV, this corresponds to a trigger threshold of 55 eV shown as the blue dotted line in Fig.~\ref{fig:Pulse} (right panel).

The triggering procedure searches for candidate events in the filtered data stream, starting with the largest positive deviation from zero. 
An exclusion interval of $\pm$0.5~s is imposed around each pulse found, and the search is reiterated in the remaining data until no fluctuation greater than $n\sigma_{OF}$ is found.
This energy ordering of the pulse finding algorithm   affects the energy dependence of the triggering efficiency. 
For instance, the dead-time associated to the search for low-energy events is effectively greater than that associated to large pulses. A dedicated data-driven procedure, described in Sec.~\ref{sect:efficiency}, has been developed to fully take into account these effects in the determination of the efficiency as a function of energy.
The resulting trigger rate is almost constant over the six days and approximately equal to 0.58~Hz, as shown in the right panel of Fig.~\ref{fig:Calibration}. 

\subsubsection{Event processing}
\label{sec:processing}

Each recorded time trace, corresponding to either triggered events or noise samples, is further processed to estimate its amplitude by minimizing the following $\chi^2$ function defined in the frequency domain as:
\begin{equation}
\chi^2(a, t_0) = \sum_i{\frac{|\tilde{v}(f_i) - a\tilde{s}(f_i)e^{-j2\pi t_0f_i}|^2}{J(f_i)}}
\label{eq:chi},
\end{equation}
where $\tilde{v}(f_i) $ is the Fourier transform of the observed signal, $a$ is the amplitude of the unit-normalized signal template $s$, and $t_0$ corresponds to the starting time of the pulse. 
For triggered events, the value of $t_0$ was allowed to vary within a 20 ms time window centered around the pulse time found by the trigger algorithm. 
In addition to delivering a slightly more precise estimation of the pulse amplitude $a$, this processing step provides a $\chi^2$ value that quantifies the quality of the fit. 
This is used to reject pulses with shapes that are not consistent with the standard pulse, such as pile-ups and other categories of spurious events further detailed in Sec.~\ref{sect:selectioncut}.

\subsection{Detector calibration and stability}

\subsubsection{Calibration}
\label{sec:calibration}

The energy calibration of the reconstructed amplitudes $a$ was ensured by the use of a low-energy X-ray $^{55}$Fe source irradiating the bottom side of the Ge crystal, opposite to the Ge-NTD heat sensor, inducing an interaction rate of $\sim$0.3 Hz. 
The $^{55}$Fe source produces two lines corresponding to the $K_\alpha$ and $K_\beta$ lines of Mn at 5.90 keV and 6.49 keV, respectively.
They are clearly visible on the left panel of Fig.~\ref{fig:Calibration}, showing the calibrated energy as a function of time over the 137-hour acquisition period. 
The energy resolution of these peaks is 34 eV (RMS). 
There is sufficient statistics to calibrate each hour separately and to follow precisely any time evolution of the detector gain, defined as the voltage sensitivity per unit of deposited energy,  and resolution. The red solid line on the same figure shows the variation of the heat sensor gain (in $\mu$V/keV)
as a function of time deduced from the peak position.
The gain fluctuations are remarkably small during the entire time period, with a mean value of 1.56~$\mu$V/keV and a maximal dispersion of $\sim 2$\%.

\begin{figure*}[!t]
\begin{center}
\includegraphics[width=0.49\textwidth,angle=0]{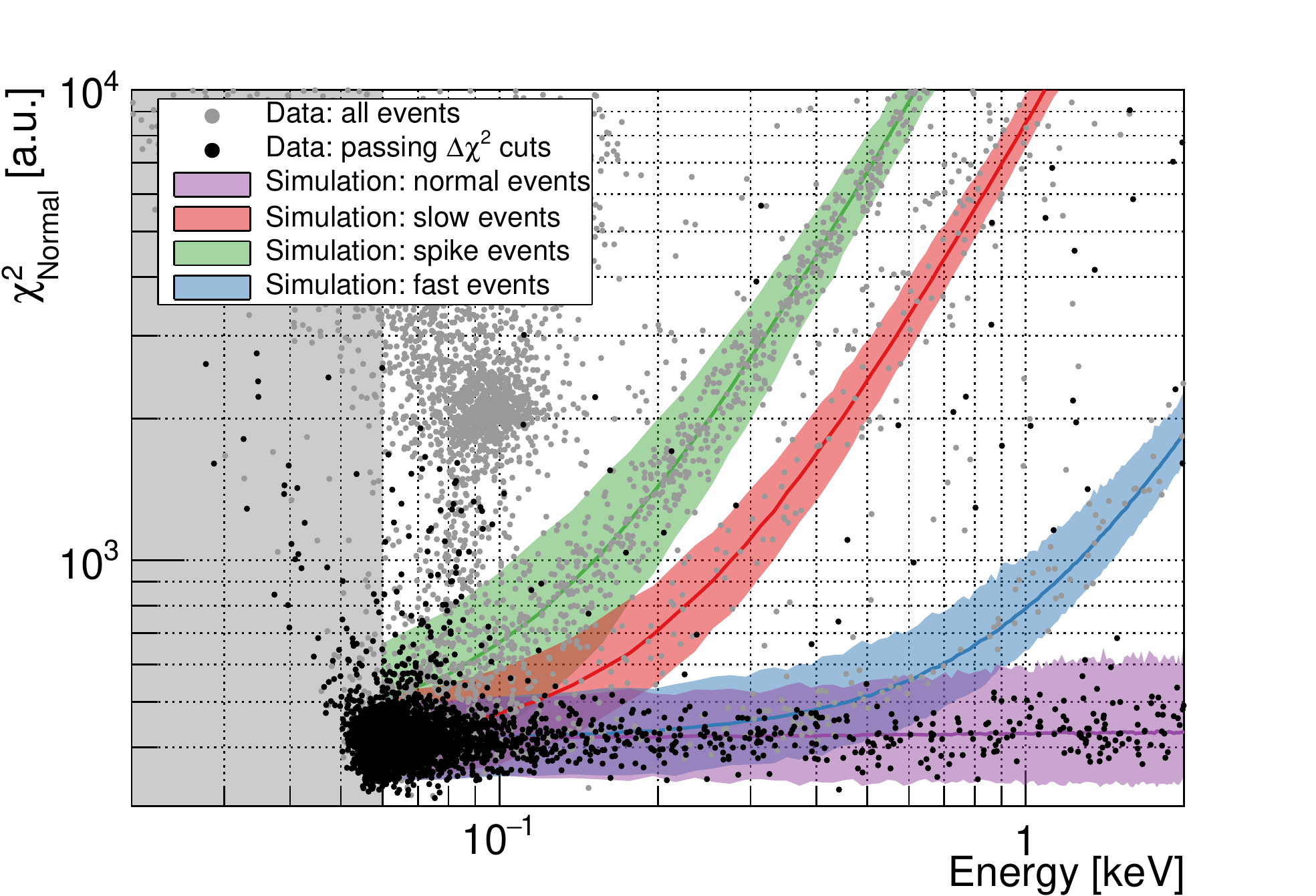}
\includegraphics[width=0.49\textwidth,angle=0]{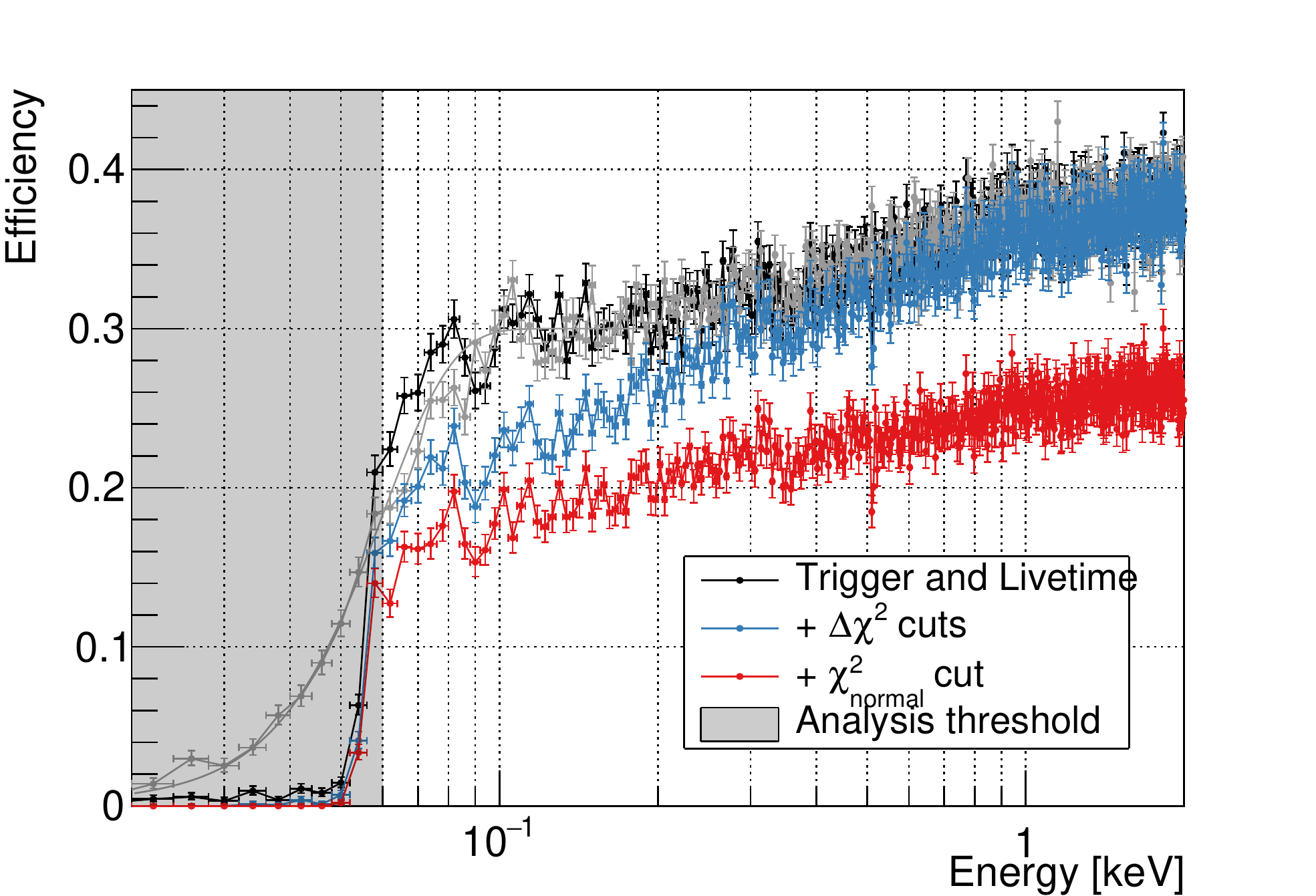}
\caption{
{\bf Left}: Distribution of $\chi^2_{\text{Normal}}$ as a function of energy for the events in the blinded data set. The events passing all $\Delta\chi^2$ cuts are shown as black dots while those rejected are shown as grey dots. The expected region where different pulse populations are expected are shown in purple (events with normal pulse shape), blue (fast pulses), red (slow pulses) and green (single-sample spikes).
{\bf Right}: Signal efficiency as a function of heat energy for $10^{6}$ simulated events injected in the data stream with energies uniformly selected between 0 and 2.5 keV, shown from 0 to 2 keV corresponding to the region of interest. Grey points: trigger plus live-time efficiency as a function of the input energy, together with a fit with an error function (grey line). Black points and line: trigger plus live-time efficiency as a function of the energy resulting from the $\chi^2$ fit of a pulse to the data in the frequency domain. 
Blue points and line: the same, but after applying all $\Delta\chi^2$ cuts.
Red points and line: the final efficiency obtained after applying the additional cut on $\chi^2_{\text{Normal}}$.
In both panels, the grey area corresponds to energies below the analysis cut.
} 
\label{fig:Trigger}
\end{center}
\end{figure*}

\subsubsection{Baseline resolution stability}
\label{sec:baseline}

As mentioned is Sec.~\ref{sec:processing}, the time traces used for the PSD determination, also called noise events, were also fitted with Eq.~\ref{eq:chi}. However, in this case  $t_0$ was arbitrarily fixed to the center of the 1-s time window, in order to estimate the amplitude distribution associated with noise events. The amplitudes are distributed according to a gaussian distribution with a standard deviation (RMS) $\sigma$ that corresponds to the baseline energy resolution.
This quantity, measured hour by hour, is compared to the prediction of the optimal filter resolution ($\sigma_{OF}$) on the right panel of Fig.~\ref{fig:Calibration}.
The good agreement between these two estimates of the energy resolution validates the assumption that the noise in each frequency bin is independent. The average baseline energy resolution is 18 eV (RMS), with a $\sim3$\% overall decrease in $\sigma_{OF}$  over the six days of the search. The average value of $\sigma$ during the blinded data period is 17.7 eV (RMS). The stability of the noise is instrumental at maintaining an almost constant trigger rate, shown as the red dots in Fig.~\ref{fig:Calibration} (right panel), over the entire search period. 

\subsection{Data analysis}

\subsubsection{Selection cuts}
\label{sect:selectioncut}
Because of its above-ground operation and the moderate lead shield, the detector was exposed to a rather intense rate of high energy interactions in the bulk of the crystal as well as in the holding PTFE clamps and the Ge-NTD heat sensor. 
The pulse shape of events produced in these cases are very different. 
In total, the following four types of events were observed: 
\begin{itemize}
\item {\em Normal} events: these correspond to the vast majority of observed events for which the incoming particle has interacted in the Ge crystal. They are characterized by a rise time of 6 ms and a decay time of 16 ms, as expected from our thermal model calculations \cite{thermModel}.
\item {\em Fast} events: these are induced by incoming particles impinging the Ge-NTD heat sensor which has a non-negligible volume of 2 mm$^3$, and from internal radioactivity of the Ge-NTD. They are characterized by a rise time of 2 ms, faster than normal events as expected, and a decay time constant of 41 ms. This longer decaying time with respect to normal events could be explained by athermal phonons heating up the Au wire bonds from the weak thermal leak. This hypothesis is further strengthened by the fact that the amplitudes of these fast events are much smaller than anticipated from our theoretical calculations \cite{thermModel}. 
\item {\em Slow} events: these could be produced by muon interactions in the holding clamps where $\mathcal{O}(1)$ MeV energy is deposited. Despite the very weak thermal coupling of these PTFE clamps to the Ge crystal, such high energy deposition can still produce some non-negligible rise in temperature of the crystal. These events are characterized by a mean rise time of 26 ms and a decay time constant of 115 ms, hence much slower than normal events. This population was later observed to be significantly reduced in detectors where the PTFE clamps have been replaced by a combination of sapphire balls with chrysocale sticks.
\item {\em Spike } events: the data acquisition system used in this experiment (a former version of the one used by EDELWEISS-III~\cite{edwtech}) suffers from random synchronization losses that result in occasional octal jumps in the data flow. These are well modelled by delta functions that can easily be discriminated from physical events.
\end{itemize}

To discriminate normal events from the other populations of events, including also pile-ups, cuts are performed on the goodness of fit parameter quantified by the $\chi^2_{\text{Normal}}$ value from Eq.~\ref{eq:chi}, where the subscript "Normal" refers to the use of the standard pulse template $s$ in that equation. 
Additional pulse shape related selection cuts have been designed based on the $\chi^2$ differences $\Delta\chi^2_k = \chi^2_{\text{Normal}} - \chi^2_k$, where $\chi^2_k$ corresponds to the value calculated by replacing the standard pulse template for normal events $s$ with the  templates of the non-standard event population discussed above, i.e. $k$ stands for {\em Fast}, {\em Slow} and {\em Spike}. 
All cuts were optimized on the non-blinded data set prior to un-blinding.

\begin{figure*}[!t]
\begin{center}
\includegraphics[width=0.9\textwidth,angle=0]{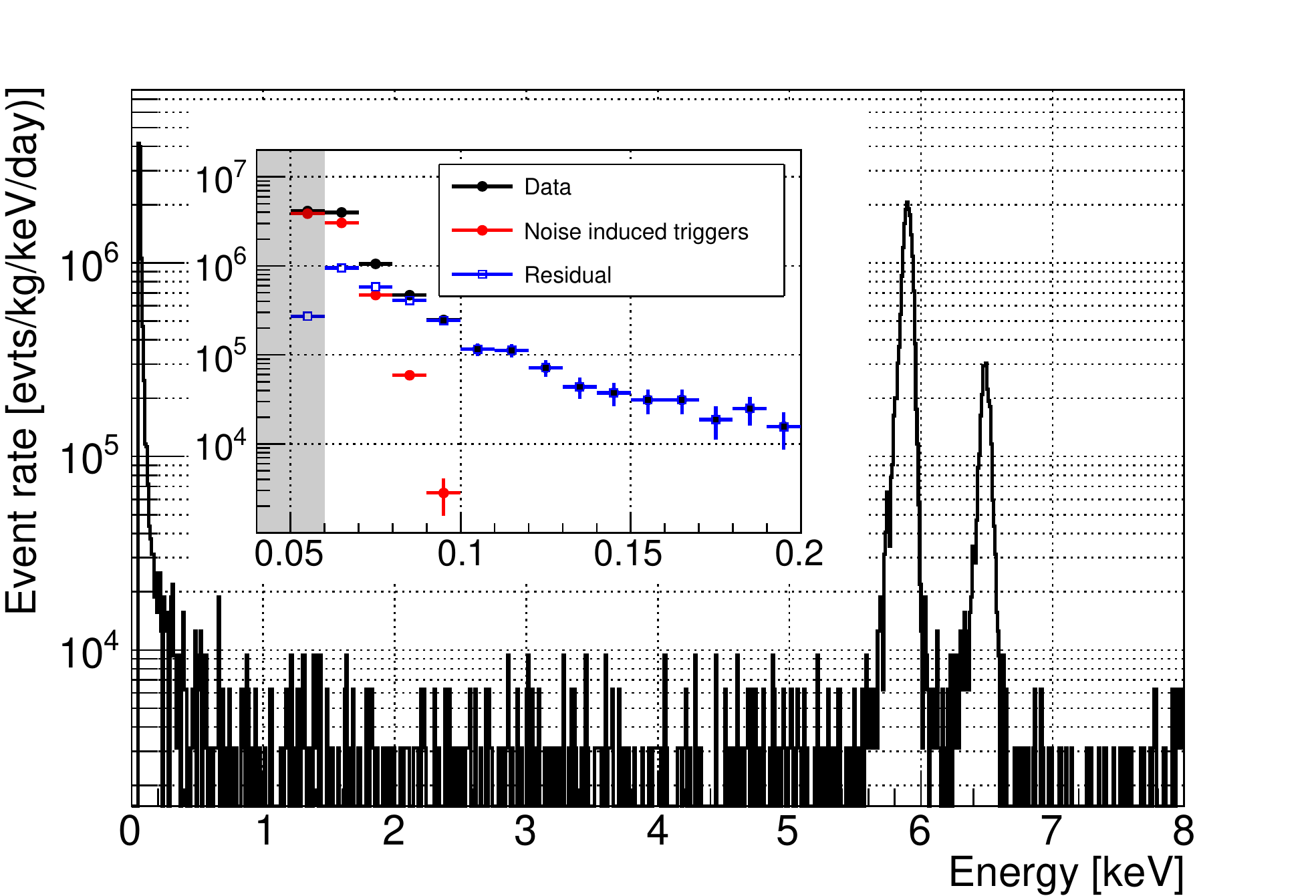}
\caption{
Energy spectrum recorded in the blinded day of the DM search data, after all cuts. The calibration lines from the $^{55}$Fe source at 5.90~keV and 6.49~keV are clearly visible, and they exhibit an energy dispersion of 34~eV (RMS). 
The data are normalized in events per kilogram per day and per keV and is not corrected for efficiency. The bin size is 10 eV.
The inset shows a zoom of this distribution between 40 eV and 200 eV (black dots), as well as the energy spectrum observed in the simulated noise streams after all cuts (red dots). The blue squares are the difference between the two spectra. 
} 
\label{fig:spectrum}
\end{center}
\end{figure*}

Figure~\ref{fig:Trigger} shows the event distribution in the $\chi^2_{\text{Normal}}$ versus energy plane before and after applying the $\Delta\chi^2$ cuts (black and grey points, respectively). The colored bands, estimated from the dedicated pulse simulation discussed in Sec.~\ref{sect:efficiency}, correspond to the 90\% C.L. confidence bands for the four different types of events. Normal events are expected to have an averaged $\chi^2_{\text{Normal}}$ value centered around $N_s = 400$ while the other event categories exhibit a quadratic divergence with increasing energy. As expected, the population of events that is the most alike to the normal ones, and therefore the hardest to reject efficiently, are the fast events. After the combination of the various $\Delta\chi^2$ cuts and selecting events with $350 < \chi^2_{\text{Normal}} < 450$ we found that the survival probabilities of the three types of spurious events reach $10^{-2}$ above 60~eV for Slow and Spike events, and above 400~eV for Fast events. 
The additional event population  observed centered at 100~eV and $\chi^2_{\text{Normal}} \sim$ 2000 corresponds to noise events where the first half of the pre-filtered time trace is affected by the tail of a previous pulse (mostly from $^{55}$Fe), occurring a few hundreds of ms before the start of its 1-s time window. Note that this particular population of pile-up events, as all the other types of pile-up events that appear in the gaps between the colored bands in the left panel of Fig.~\ref{fig:Trigger}, is fully taken into account in the pulse simulation procedure described in Sec.~\ref{sect:efficiency}, and is very well rejected by the standard $\chi^2_{\text{Normal}}$ cut.

\subsubsection{Efficiency}
\label{sect:efficiency}

The trigger efficiency was evaluated using a dedicated pulse simulation where pulses are generated at random times throughout the entire real data streams. 
This procedure samples rigorously all possible baseline fluctuations, including those induced by tails of high-energy events, and other non-stationary behaviour, hence avoiding any possible selection-induced biases. 
It naturally accounts for live-time losses due to the physical event rate and comprehensively takes into account any systematic uncertainties or biases related to the processing pipeline. 
The following efficiency estimates, as well as the various simulated DM signals discussed in Sec.~\ref{sec:results}, were obtained by generating a total of 10$^6$ simulated events distributed over 1000 Monte Carlo iterations of injecting 1000 simulated pulses in the 24-hour long blinded data stream. 
This way, the simulated event rate is about two orders of magnitude lower than the real physical event rate, hence inducing negligible additional live-time losses.
A full simulation of the DM signal is performed for each mass value in order to evaluate in each case the combined effect of the noise observed in the actual data stream, the trigger selection and the analysis cuts. To better illustrate and understand the detector performance, it is however useful to calculate an efficiency taken as the fraction of simulated events that survive all of these selection criteria, from a population of simulated events with energies uniformly distributed between 0 and 2.5 keV.
These resulting efficiencies at different stages of the trigger and analysis are shown in the right panel of Fig.~\ref{fig:Trigger}. 
The black and grey points correspond to the variation of the combined trigger and live-time efficiency as a function of the reconstructed and input energy, respectively. 
The sharp rise around 55 eV of the efficiency curves expressed as a function of the reconstructed energy is due to the fact that the amplitude estimates from the triggering algorithm and the pulse processing are very similar. 
The rise of the efficiency as a function of the true input energy is much smoother and is well described by an error function centered at 55 eV with a dispersion of 18 eV, as anticipated from the observed baseline energy resolution and the considered 3$\sigma$ trigger level.
As discussed in Sec.~\ref{sec:trigger}, the trigger efficiency rises smoothly with energy, because of the energy ordering of the algorithm. 
Hence, between 0.1 and 2.0 keV, the trigger and live-time efficiency rises from 30\% to almost 40\%.
These large dead-time losses can be expected given the size of the $\pm0.5$~s exclusion time window compared to the observed trigger rate of 0.58~Hz.

It has been tested that the dead time could be reduced by using shorter time traces, but at the cost of a slight deterioration of the energy resolution. 
As it was anticipated that this high-background DM search would not be limited by statistics, the emphasis was put on low thresholds and therefore energy resolution.

The right panel of Fig.~\ref{fig:Trigger} also shows the effect of the $\Delta\chi^2$ on the analysis efficiency. Those cuts have a significant effect on the signal below 200 eV, where all three background populations compete with the signal.
Finally, the cut on the $\chi^2_{\text{Normal}}$ value has a more uniform effect as a function of energy, as shown as the red points on the same figure.

\begin{figure*}[!t]
\begin{center}
\includegraphics[width=0.49\textwidth,angle=0]{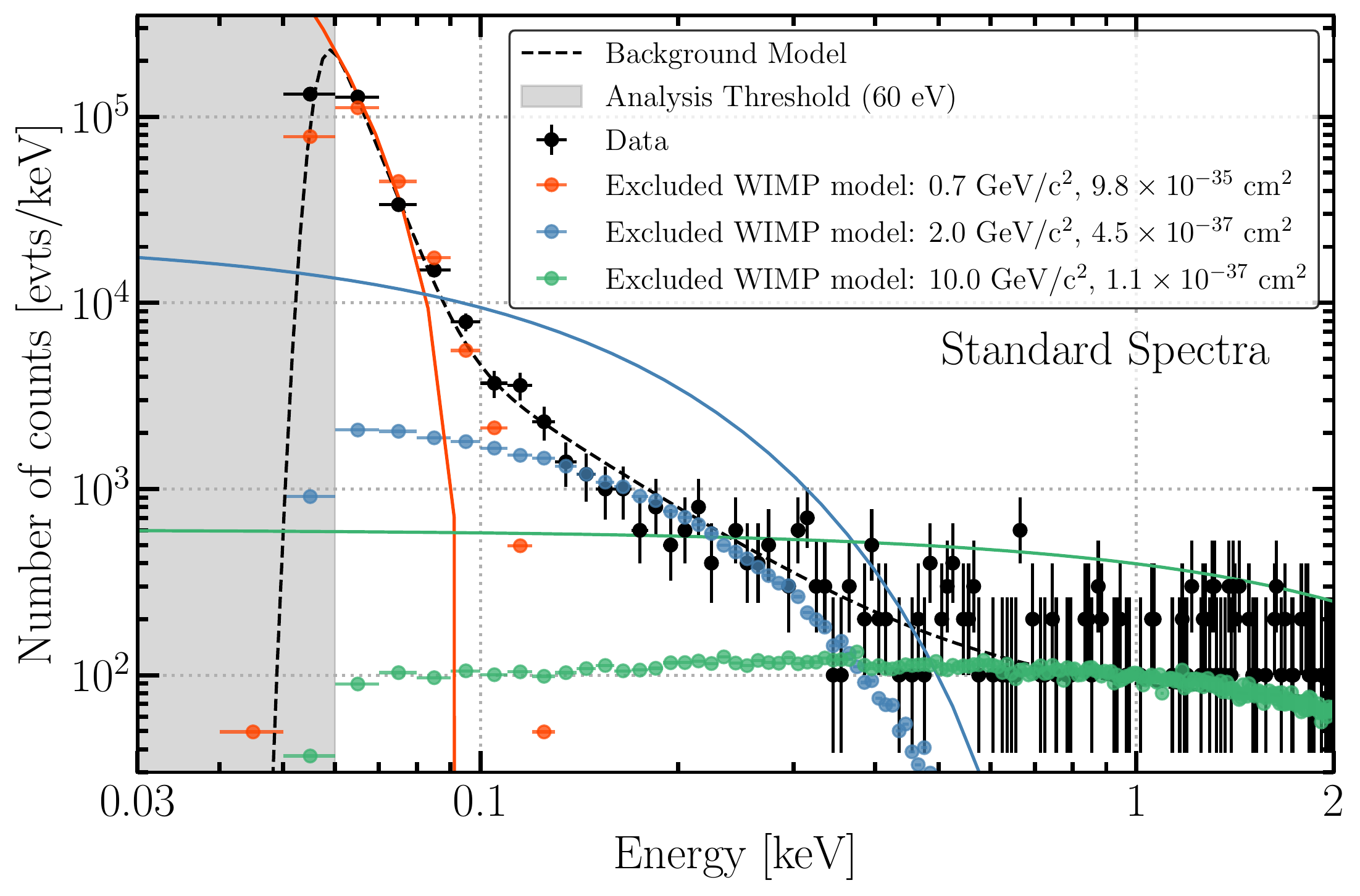}
\includegraphics[width=0.49\textwidth,angle=0]{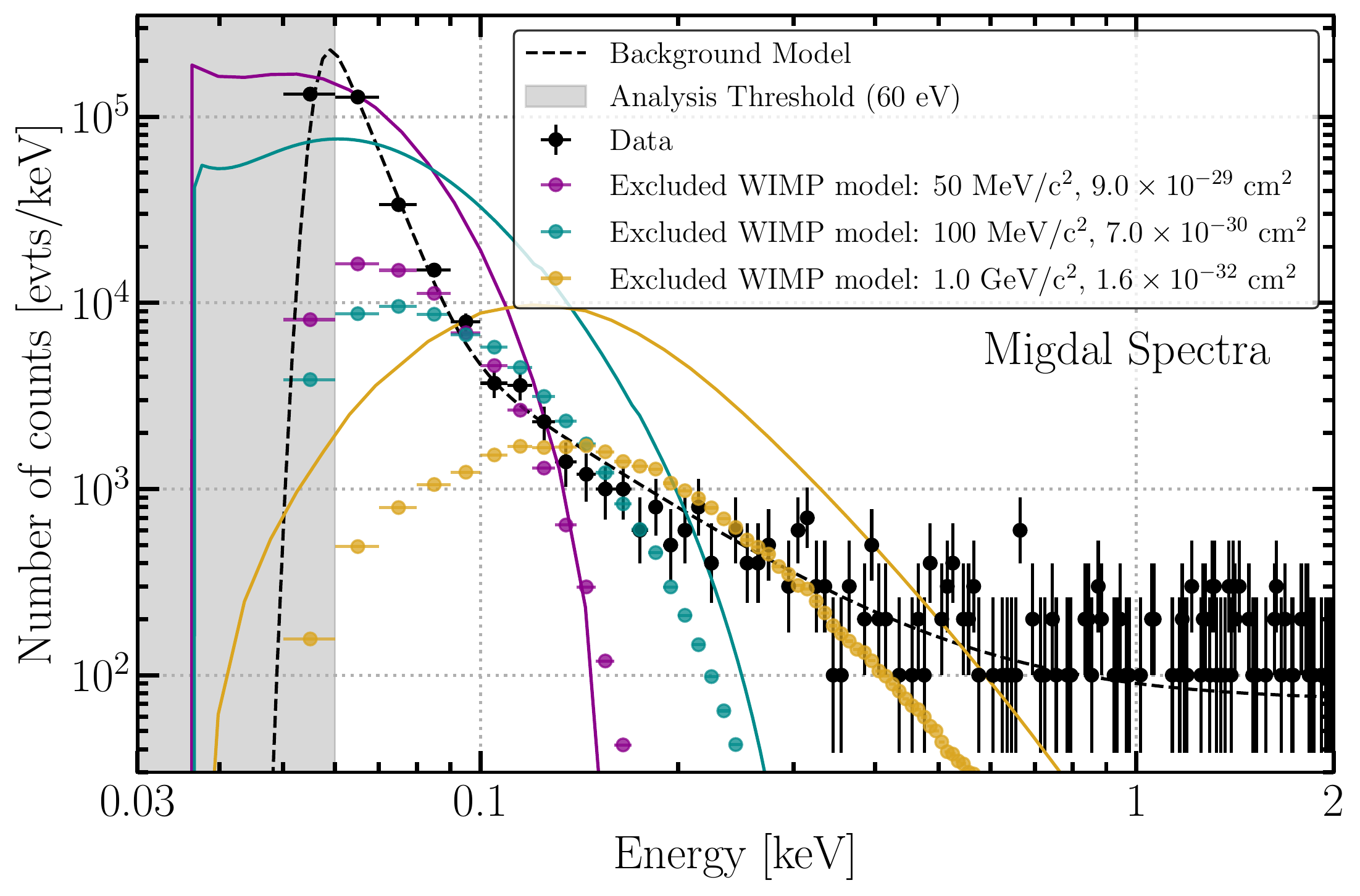}
\caption{Comparison of the energy spectrum after all cuts for the blinded data set (black dots) with the background model derived from the five-day non-blinded data set (black dashed line). Also shown are the spectra for three excluded WIMP signals. 
The solid colored lines are the unsmeared WIMP spectra, while colored points are the spectra that incorporate resolution effects and all cuts.
{\bf Left}: Standard elastic nuclear recoil spectra that are excluded for WIMPs with masses of 0.7 GeV/c$^{2}$ (red), 2 GeV/c$^{2}$ (blue) and 10 GeV/c$^{2}$ (green). 
{\bf Right}: Inelastic Migdal spectra that are excluded for WIMPs with masses of 0.05 GeV/c$^{2}$ (magenta), 0.1 GeV/c$^{2}$ (cyan) and 1.0 GeV/c$^{2}$ (yellow).} 
\label{fig:signals}
\end{center}
\end{figure*}

\subsubsection{Observed energy spectrum}

The energy spectrum recorded within the blinded day of DM search data is shown in Fig.~\ref{fig:spectrum}.
The spectrum is dominated at high energy by the calibration source lines at 5.90 keV and 6.49 keV. 
A continuous and relatively flat background of $8\times10^3$ evt/kg/keV/day is observed between 500 eV and 8 keV\footnote{The quoted background rates are corrected for the final efficiency as a function of energy (see red curve in Fig.~\protect{\ref{fig:Trigger}}).}. 
The data below 500 eV, better illustrated in the inset figure, can be described approximately by an exponentially decreasing spectrum with a characteristic slope of 25 eV reaching $10^5$ evt/kg/keV/day at 200 eV and culminating at $2\times10^7$ evt/kg/keV/day at the 60~eV analysis energy threshold. While the flat component of $8\times10^3$ evt/kg/keV/day corresponds to the expected gamma-ray background for a detector operated in a non-low-radioactivity cryostat with a moderate lead shield, the exponential rise at the lowest energies is not yet fully explained. Dedicated studies are ongoing to better assess their origin, such as adding electrodes to read out the ionization yield associated to these interactions, and modifying the detector holders. 

Finally, the inset figure also shows the contribution from the noise induced triggers (red histogram) that has been evaluated using the noise-only simulated data stream discussed in Sec.~\ref{sec:trigger}. 
This event population explains very well the observed spectrum below 70~eV and it has a negligible contribution to the total observed event rate above 80~eV. 
It should be noted that in order to derive conservative constraints on DM interactions, it was decided prior to un-blinding that this well anticipated and modeled background will not be subtracted.

\section{Results}
\label{sec:results}

This section presents the limits on spin-independent interactions of DM particles with nucleons derived from the experimental spectrum of Fig.~\ref{fig:spectrum}.
In Sec.~\ref{sec:WIMP} the data are interpreted in the standard WIMP context where the signal is the kinetic energy of the nuclear recoil induced in the collision. In Sec.~\ref{sec:SIMP} we extend its interpretation to the case of SIMPs where Earth-shielding effects are taken into account. In Sec.~\ref{sec:Migdal} we consider the so-called Migdal effect where the DM-nucleus interaction is inelastic and simultaneously produces a nuclear and an electronic recoil, opening up access to an unexplored domain of cross sections for DM particles with masses well below 100~MeV/c$^{2}$.
Finally, these results are interpreted in terms of spin-dependent interactions of DM particles in Sec.~\ref{sec:SDlimits}.

\subsection{Weakly Interacting Spin-independent Dark Matter}
\label{sec:WIMP}
The expected signal for the standard elastic DM interactions in the detector was simulated using the comprehensive pulse simulation, described in Sec.~\ref{sect:efficiency}, that takes into account all systematic effects related to the detector response and data analysis. A total of $10^6$ fake DM-induced events were simulated for each DM mass considered.
Their amplitudes were drawn from the theoretical distribution of induced nuclear recoils calculated assuming standard spin-independent interactions and using the standard astrophysical parameters for the DM velocity distribution~\cite{standardassumptions}: a Maxwellian velocity distribution with an asymptotic velocity of $v_0 = $ ~220~km/s, and a galactic escape velocity of $v_{\text{esc}} = $~544~km/s. The local DM density at Earth position is assumed to be $\rho_0 = $~0.3~GeV/c$^{2}$/cm$^3$ and the lab velocity with respect to the DM halo is $V_{\text{lab}} = $ 232 km/s. Also, the loss of coherence at high momentum is taken into account using the standard Helm form factor.

The objective of the search was to establish upper limits on the DM-nucleon cross section for each considered DM particle mass, using Poisson statistics and assuming that all events observed in a given energy interval are signal candidates.
Prior to un-blinding the DM search data, the optimal energy intervals for such a purpose were determined by maximizing for each DM particle mass the signal-to-noise ratio between the simulated signals and the background model extracted from the 5 days of non-blinded data\footnote{In the presence of a signal, this procedure overestimates the background and yields conservative bounds.}.

After the un-blinding, the numbers of counts in these intervals were extracted from the data shown in Fig.~\ref{fig:spectrum}.
In the left panel of Fig.~\ref{fig:signals},
the derived 90\% C.L. upper limits of signals for WIMPs with masses of 0.7, 2 and 10 GeV/c$^{2}$ are compared to the experimental data. 
The background model, i.e. the average spectrum recorded in the 5 non-blinded days, is also shown. 

Finally, the spin-independent WIMP-nucleon cross sections excluded at 90\% C.L.\ as a function of the WIMP mass are shown as the solid red curve in Fig.~\ref{fig:Limit}, and compared with the other experimental results from Refs.~\cite{xenon1t,lux,cdmslite,cresst,edwiii,darkside,nucleus,newsg,xqc,Dolan:2017xbu,Akerib:2018hck}, and the so-called neutrino floor~\cite{neutrinoBillard}.
The EDELWEISS-Surf result is the most stringent, nuclear recoil based, above-ground limit on spin-independent interactions above 600~MeV/c$^{2}$.

\subsection{Strongly Interacting Dark Matter}
\label{sec:SIMP}

\begin{figure*}[!t]
\begin{center}
\includegraphics[width=0.9\textwidth,angle=0]{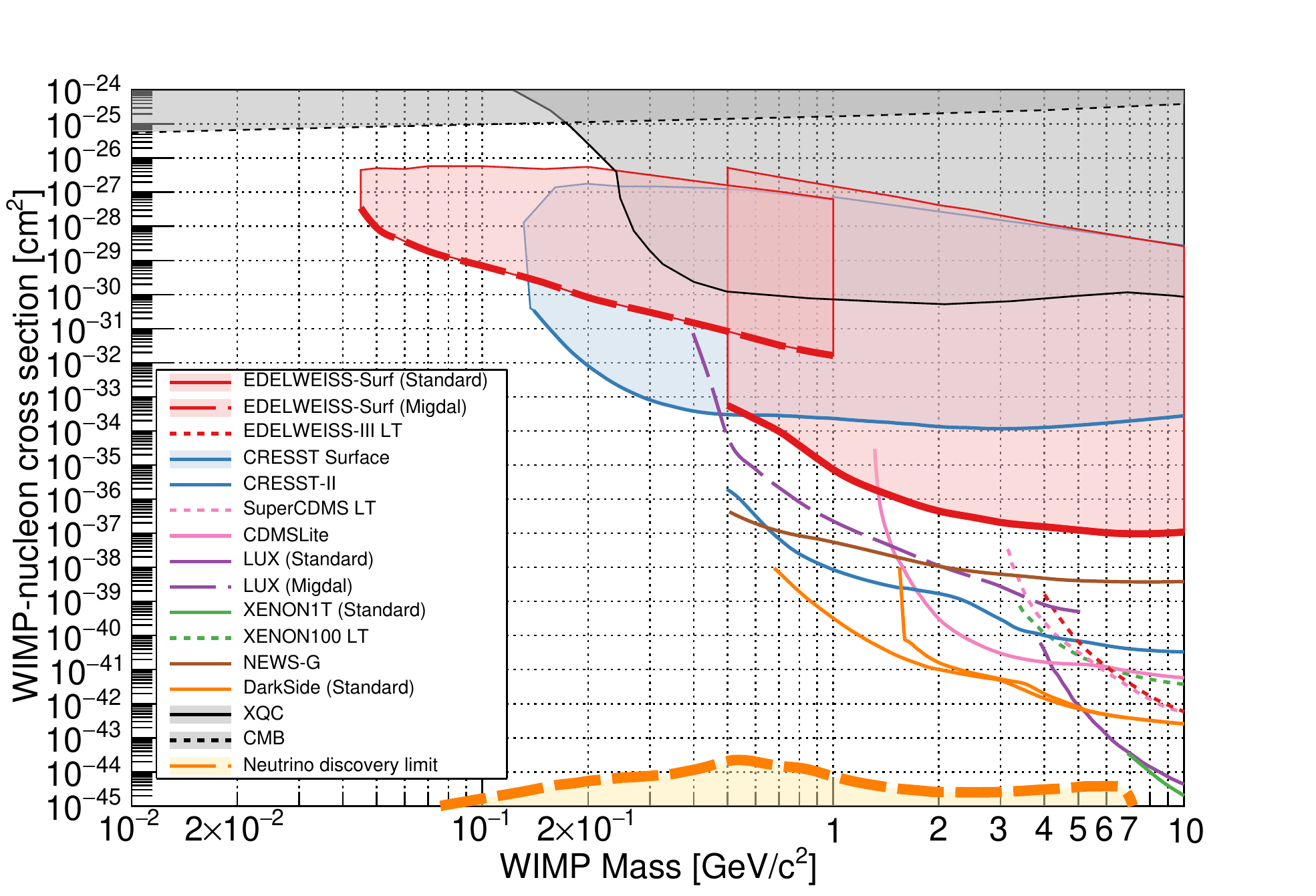}
\caption{The 90\% C.L. limits on the cross section for spin-independent interaction between a DM particle and a nucleon as a function of the particle mass obtained in the present work. The thick solid red line corresponds to the result from the standard WIMP analysis. The associated red contour is obtained from the SIMP analysis, taking into account the slowing of the DM particle flux through the material above the detector. The thick dashed line and its accompanying red contour is obtained in the Migdal analysis. These results are compared to those of other experiments (see text). Other results using the Midgal effect are shown as dashed lines. The other shaded contours correspond to the SIMP analyses of the CRESST 2017 Surface Run \cite{nucleus,Davis:2017noy,Kavanagh:2017cru} (blue contour), the XQC rocket~\protect{\cite{xqc,Mahdawi:2018euy}} (grey contour with full line) and the CMB~\cite{Gluscevic:2017ywp} (grey contour with dashed line). }
\label{fig:Limit}
\end{center}
\end{figure*} 

Thanks to its above-ground operation, the present DM search can probe SIMPs that would escape detection in underground experiments as the DM particles would be stopped in the rock overburden before reaching the detectors. 
We therefore extend the data interpretation of Sec.~\ref{sec:WIMP} to determine for each mass a range of excluded cross sections that take into account the absorption of the SIMP flux in the material above the detector. 
Overly conservative SIMP limits can be obtained by including in the analysis only those SIMPs that reach the detector without scattering \cite{Hooper:2018bfw}. 
Here, more stringent limits are obtained by fully taking into account the effect of scattering on the velocity distribution of the SIMP flux reaching the detector.
This velocity distribution is calculated using the publicly-available \texttt{verne} code~\cite{verne}, introduced in Ref.~\cite{Kavanagh:2017cru}. 
It assumes continuous energy losses and straight-line trajectories for SIMPs travelling through the atmosphere, overburden and detector shielding \cite{Starkman:1990nj}. Ref.~\cite{Emken:2018run} showed that this simplified formalism leads to constraints similar to more complete but computationally expensive Monte Carlo simulations (e.g.~\cite{Emken:2017qmp,Mahdawi:2017cxz,Mahdawi:2017utm,Mahdawi:2018euy}). 

The SIMP flux calculation takes into account the variation of the direction of the mean DM flux \cite{Mayet:2016zxu} over the course of the blinded EDELWEISS-Surf exposure (24 hours, starting 17h00 on 26th May 2018). 
It also accounts for the atmospheric overburden above the detector as well as shielding provided by the material in the building where the detector is located and the lead, steel and copper components in its vicinity. 
The dominant sources of stopping for downward-travelling particles are 20 cm (40 cm) of concrete from the ceiling (walls) and 10 cm of lead shielding which surrounds the detector in all directions, apart from an opening of around $50^\circ$ above the detector. 
Upward-travelling particles are almost entirely stopped by the Earth.

The escape velocity from the surface of the Earth is around 11~km/s. 
For DM particles at such low speeds, effects such as gravitational capture \cite{Gould:1987ww,Mack:2007xj} and gravitational focusing \cite{Kouvaris:2015xga} may become important. 
These effects are not incorporated in the flux calculation. 
Instead, the DM velocity distribution is conservatively set to zero below $v_\mathrm{cut} = 20$~km/s when calculating SIMP bounds.

Because of the very large values of cross sections involved and consequently large attenuation of the flux, the simulation of the SIMP signals corresponding to the upper bound of the excluded cross section contour requires samples many orders of magnitude larger than those required in the simple WIMP analysis of Sec.~\ref{sec:WIMP}.
As scaling up the simulated sample size from $10^6$ to $>10^{10}$ was not technically feasible for computational reasons, we developed an analytic model for the detector response based on the simulation of $10^7$ events with input energies ranging from 0 to 2.5 keV (see Appendix~\ref{sec:response}).
This model describes the probability $P_\mathrm{OF}(E_\mathrm{out}|E_\mathrm{in})$ of reconstructing an energy $E_\mathrm{out}$ given an initial energy $E_\mathrm{in}$ when applying the optimal filter algorithm of Sec.~\ref{sec:processing}.
The observed spectrum of events $\frac{\mathrm{d}R}{\mathrm{d}E_\mathrm{out}} $ is thus given by:
\begin{align}
    \frac{\mathrm{d}R}{\mathrm{d}E_\mathrm{out}} = \eta(E_\mathrm{out})\int_0^\infty  P_\mathrm{OF}(E_\mathrm{out}|E_\mathrm{in}) \frac{\mathrm{d}R} {\mathrm{d}E_\mathrm{in}}\,\mathrm{d}E_\mathrm{in}\,.
\end{align}
The measured efficiency as a function of output energy is $\eta(E_\mathrm{out})$, as shown by the red curve in the right panel of Fig.~\ref{fig:Trigger}. 
The calculation of $P_\mathrm{OF}$ and the comparison of the analytic detector response with results of the pulse simulations is discussed in Appendix~\ref{sec:response}.

Using the signal calculated in these  simulations, the same statistical procedure described in Sec.~\ref{sec:WIMP} is applied to derive the 90\% C.L. upper bounds on the excluded cross section interval as a function of SIMP mass, resulting in the red contours shown in Fig.~\ref{fig:Limit}. The upper bound reported in this work improves upon the high-cross section reach of the CRESST 2017 surface run \cite{nucleus} (thin blue), as reported in Refs.~\cite{Davis:2017noy, Kavanagh:2017cru,Emken:2018run}. This improvement is driven in part by the longer exposure of the EDELWEISS-Surf run, which covers a full day. This includes periods when the mean direction of the DM flux (set by the Sun's velocity) is perpendicular to the Earth's surface, reducing the attenuation effect of the Earth and atmosphere. 

\begin{figure*}[!t]
\begin{center}
\includegraphics[width=0.49\textwidth,angle=0]{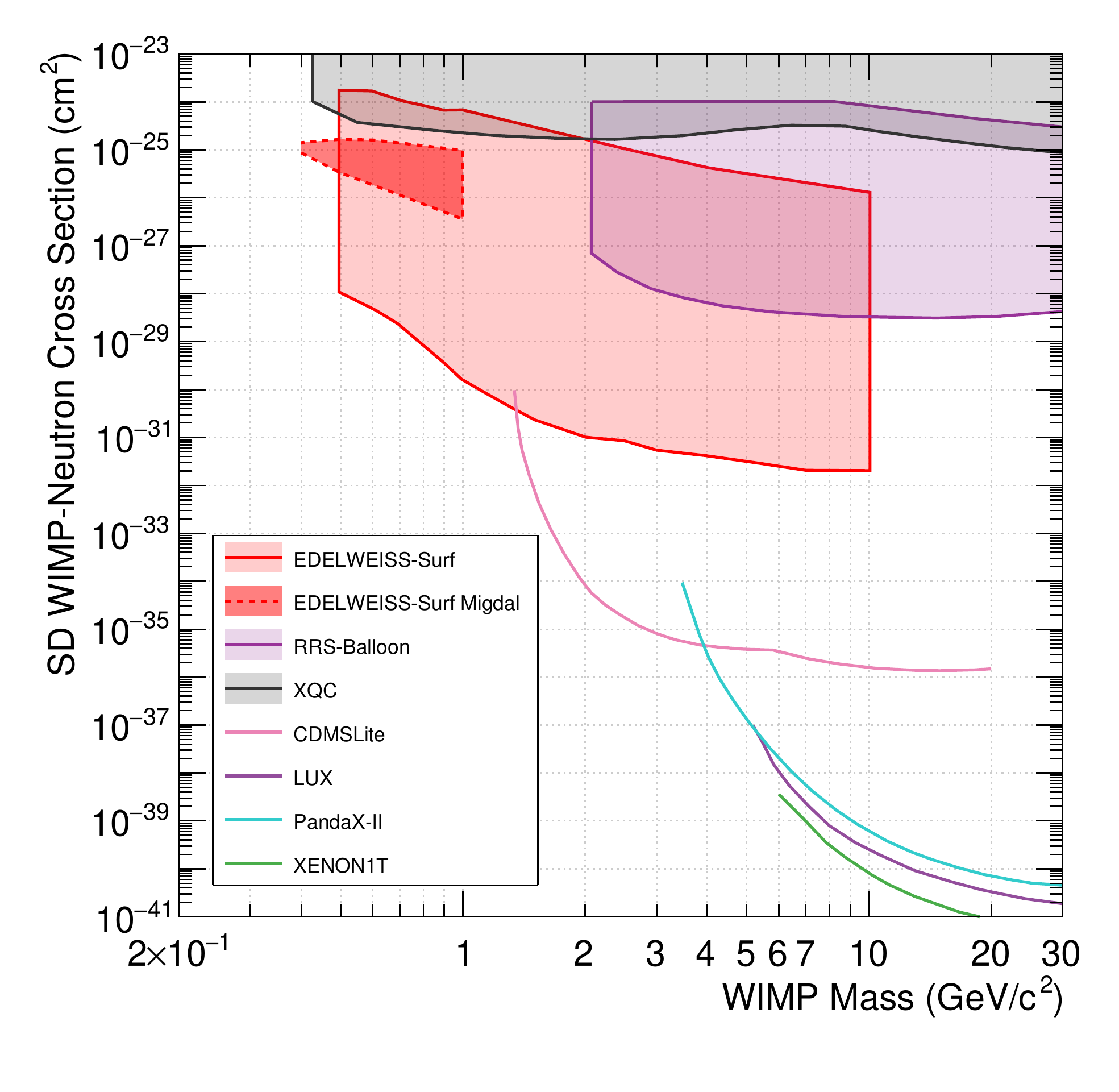}
\includegraphics[width=0.49\textwidth,angle=0]{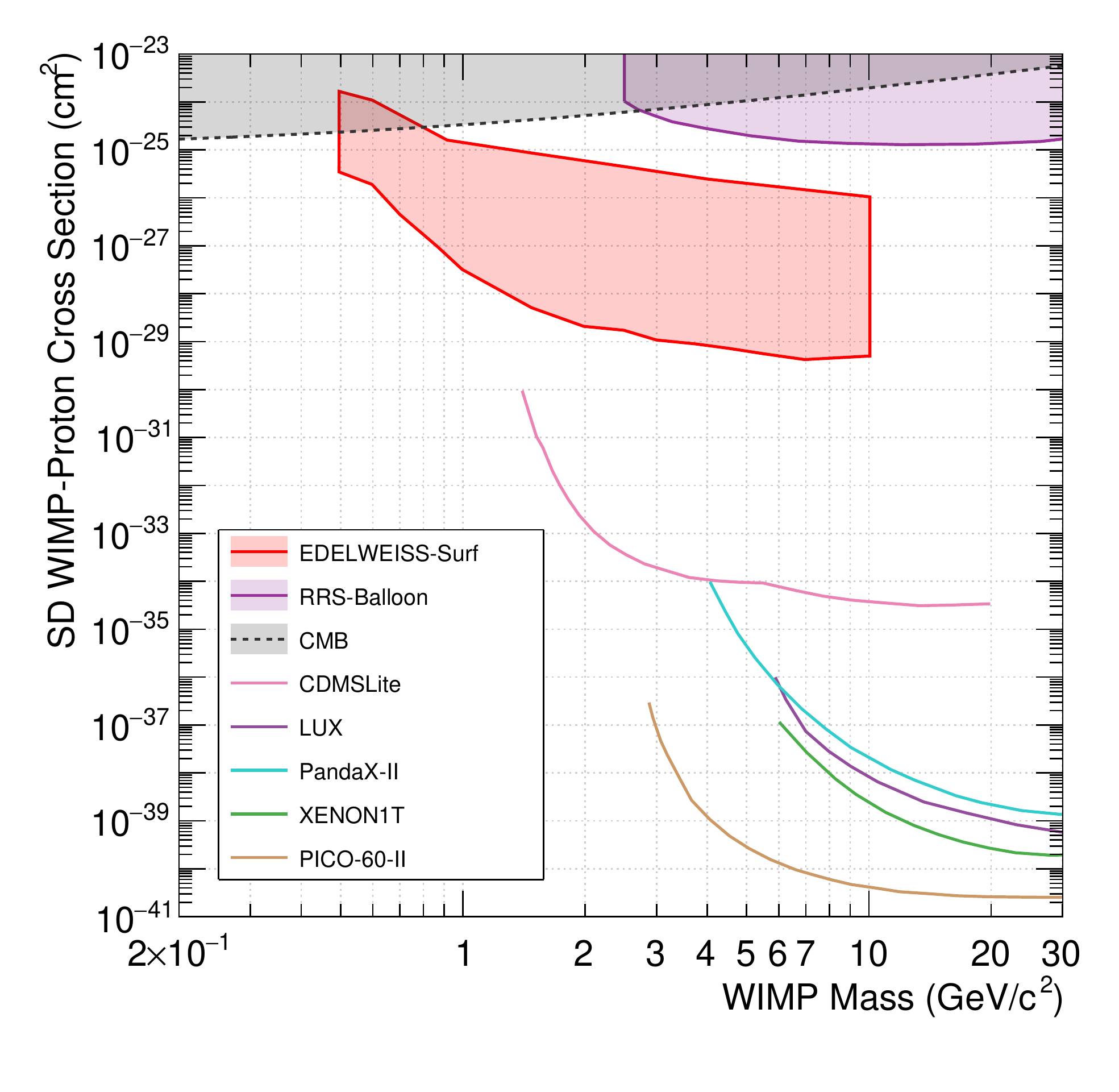}
\vspace{-0.5cm}
\caption{The 90\% C.L. limits on the cross section for spin-dependent interactions assuming a dark matter coupling only to neutrons (left panel) and to protons (right panel) as a function of the particle mass obtained in the present work. The thick red lines and contours correspond to the presented surface dark matter search taking into account Earth-shielding effects and the so-called Migdal effect (dashed line) which is only relevant for the neutron SD coupling. These results are compared to those of other direct detection experiments shown as solid lines: LUX~\cite{Akerib:2017kat} (purple), XENON1T~\cite{Aprile:2019dbj} (green), PICO-60-II~\cite{Amole:2019fdf} (brown), CDMSLite~\cite{Agnese:2017jvy} (pink), and PANDAX-II~\cite{Fu:2016ega} (blue).  The other shaded contours correspond to the SIMP analyses from the XQC rocket~\protect{\cite{xqc,Mahdawi:2018euy,Hooper:2018bfw}} (black solid line), the RRS baloon experiment~\cite{Hooper:2018bfw}, and the CMB~\cite{Gluscevic:2017ywp} (grey contour with dashed line).}
\label{fig:LimitSD}
\end{center}
\end{figure*} 

Sensitivities to 100~MeV/c$^2$-scale SIMPs with spin-independent cross section of the order of 10$^{-30}$ cm$^2$ have been reported in~\cite{Collar:2018ydf}. These were derived based on the dark counts observed in a photo-multiplier tube coupled to a liquid scintillator cell operated in a shallow site with 6 meter water equivalent overburden. However these resulting constraints did not take into account Earth-shielding effects, which are particularly relevant for such low-mass dark matter particles with cross section values above $10^{-31}$~cm$^2$, hence preventing them to be compared with the ones presented in this work.

\subsection{Migdal Search}
\label{sec:Migdal}

As discussed in Sec.~\ref{sec:detector}, the detector acts as a true calorimeter with equal sensitivity to the energy deposited by nuclear and electronic recoils. 
In this section, we consider the case where the WIMP or SIMP interaction with the target atoms induces simultaneously a nuclear recoil and the ionization of an electron.
The final state comprises the two types of recoils.
This is true in the case of the so-called Migdal effect, whose rate was calculated numerically in Ref.~\cite{Ibe:2017yqa}. The calculations therein were performed for the case of \textit{isolated} atoms and does not fully consider the band structure of the germanium semi-conductor which is particularly important for the valence electrons ($n=4$). 

In the absence of detailed calculations, we chose to neglect the contribution from the ionization of electrons in the $n=4$ valence shell, as well as the much smaller contribution from excitation of electrons \textit{into} the valence shell. 
The inner electrons $n \leq 2$ are too tightly bound to give an appreciable signal.
Therefore, the only contribution considered here is that from the ionization of M-shell ($n = 3$) electrons. 
Once ionized, electrons are not free (as in Ref.~\cite{Ibe:2017yqa}), but instead populate the conduction band. 
However, the band gap in germanium is typically much smaller than the M-shell ionization energies ($\sim 0.74 \,\mathrm{eV}$ compared to $35\text{--}170\,\mathrm{eV}$, respectively) and here we will also neglect this small correction. 
Electrons, radiation and nuclear recoils in that energy range have very short absorption lengths in germanium, and it can be considered that the detector will completely collect the energy from all contributions.

The same standard spin-independent DM-nucleus interactions are assumed as in the previous sections, and notably the ($\propto A^2$) dependence of the cross section arising from the coherent coupling to the whole nucleus. 
The Earth-shielding effects are taken into account as in Sec.~\ref{sec:SIMP}. 
While the observable signal arises from inelastic Migdal events, the predominant stopping power comes from the elastic DM-nucleus collisions, and therefore only these are taken into account in the calculation of Earth-shielding effects. 

The right panel of Fig.~\ref{fig:signals} shows the expected Migdal spectra for WIMPs with masses of 0.05, 0.1 and 1.0 GeV/c$^{2}$ corresponding to the respective interaction cross section values excluded at 90\% C.L.
The comparison with the standard signal in the left panel for WIMPs with considerably larger masses clearly shows how the Migdal effect greatly enhances the sensitivity to very light WIMPs, albeit for significantly larger cross section values.

 The resulting 90\% C.L.\ exclusion region is shown in Fig.~\ref{fig:Limit}, bounded from below by the thick dashed red line. 
 The constraints weaken rapidly for WIMPs with masses below 100~MeV/c$^{2}$. 
 For lighter WIMPs, the peak in the expected Migdal signal spectrum overlaps in large part with the observed rapid rise of the data below 100~eV. 
 Nevertheless, limits are obtained for WIMP masses as low as 45~MeV/c$^{2}$. 
 Below this, the large cross sections required to give an observable signal lead to substantial stopping effects in the atmosphere and shielding, meaning that no constraints can be obtained anymore.

For comparison, we also show in Fig.~\ref{fig:Limit} the Migdal limit obtained by the LUX collaboration \cite{Akerib:2018hck} (purple dashed line) but we do not show the unofficial limits based on XENON1T as presented in Ref.~\cite{Dolan:2017xbu}.

\subsection{Spin-dependent Interactions}
\label{sec:SDlimits}

In this last section we present our result in the context of a spin-dependent coupling between the dark matter particle and the nucleus. The isotope $^{73}\mathrm{Ge}$ (with a natural abundance of 7.73\%) is the only stable germanium isotope with non-zero nuclear spin ($J = 9/2$), hence providing sensitivity to spin-dependent (SD) WIMP-nucleus interactions. For the spin expectation values, we assume $\langle S_p\rangle = 0.031$ and $\langle S_n\rangle = 0.439$ \cite{Klos:2013rwa}, with the larger $\langle S_n \rangle$ value is coming from a single unpaired neutron. For the spin-structure functions, we take the average of the values reported in Table~V of Ref.~\cite{Klos:2013rwa} which includes both 1- and 2-body currents.

We account for Earth-scattering effects, including only the SD scattering of the dark matter particles with nitrogen-14 in the atmosphere, which is the dominant source of stopping for surface-based SD searches \cite{Hooper:2018bfw}. Nitrogen-14 has nuclear spin $J=1$ and we assume equal proton and neutron spin contents $\langle S_p\rangle = \langle S_n\rangle = 1$. The results, presented as 90\% C.L. excluded regions and limits, are shown in Fig.~\ref{fig:LimitSD}  for DM-neutron and DM-proton SD couplings on the left and right panels respectively. Our results, show as red lines and contours, are compared to those of other direct detection experiments shown as solid lines: LUX~\cite{Akerib:2017kat} (purple), XENON1T~\cite{Aprile:2019dbj} (green), PICO-60-II~\cite{Amole:2019fdf} (brown), CDMSLite~\cite{Agnese:2017jvy} (pink), and PANDAX-II~\cite{Fu:2016ega} (blue).  The other shaded contours correspond to the SIMP analyses from the XQC rocket~\protect{\cite{xqc,Mahdawi:2018euy,Hooper:2018bfw}} (black solid line), the RRS baloon experiment~\cite{Hooper:2018bfw}, and the CMB~\cite{Gluscevic:2017ywp} (grey contour with dashed line). As for the SI case, see Sec.~\ref{sec:SIMP}, the above-ground dark matter searches reported in~\cite{Collar:2018ydf, Abdelhameed:2019szb} do not take into account Earth-shielding effects which are particularly relevant for cross section values above $10^{-31}$~cm$^2$, and are therefore not shown in Fig.~\ref{fig:LimitSD}.

The EDELWEISS-Surf results are exploring new parameter space of the SD couplings, on both neutrons and protons, for dark matter masses ranging from 500 MeV/c$^2$ to  1.3~GeV/c$^2$ considering the standard DM induced nuclear recoil signal. The Migdal effect, producing both a nuclear recoil and an ionised electron, extends our lower DM mass bound down to 400 MeV/c$^2$ only in the SD-neutron case because of the dominating Earth-shielding effects happening at such high cross sections. In the case of WIMP-proton SD interactions, we obtain no limit from the Migdal effect due to the strong Earth-shielding effects at such high cross sections. Indeed, the combination of the small proton spin content of $^{73}\mathrm{Ge}$ and the low ionization probability means that the scattering rate in this case is extremely small, hence requiring a large cross section to be detectable (around $10^{-24}\,\mathrm{cm}^2$ at $1 \,\mathrm{GeV}/c^2$). However, at such large cross sections, atmospheric stopping is too strong for a DM particle to reach our detector. We also note that at low masses ($\mathcal{O}(500 \,\mathrm{MeV}/c^2)$ both the Migdal WIMP-neutron and the standard WIMP-proton constraints are affected by Earth-stopping effects at both the upper and lower parts of the excluded regions. 

\section{Conclusion}
\label{sec:conclusion}

The EDELWEISS collaboration has searched for dark matter particles with masses between 45~MeV/c$^{2}$ and 10~GeV/c$^{2}$ using a 33.4-g germanium detector operated in a surface lab, and thus relevant in the search for Strongly Interacting Massive Particles (SIMPs).
The energy deposits were measured using a Ge-NTD thermal sensor with a 17.7~eV (RMS) heat energy resolution leading to a 60~eV analysis threshold. 
This performance, combined with the nearly completely stationary behavior of the detector, led to the achievement of the first limit for the spin-independent interaction of sub-GeV WIMPs based on a germanium target. 
The experiment provides the most stringent, nuclear recoil based, above-ground limit on spin-independent interactions above 600~MeV/c$^{2}$. 

The search results were also interpreted in the context of SIMPs, taking into account the screening effect of the atmosphere and material surrounding the detector.
The lower part of the excluded region for SIMPs corresponds to the previously mentioned WIMP limit and represent the most stringent constraint for masses above 600~MeV/c$^{2}$.
The upper part of the excluded region is limited by Earth-shielding effects: it probes the largest SIMP-nucleon cross sections of any direct detection experiment, excluding a value of $10^{-27}\,\mathrm{cm}^2$ for a 1 GeV/c$^{2}$ WIMP\footnote{Note that CDMS-II provides the strongest SIMP constraints for DM masses above around $10^8$ GeV/c$^{2}$~\cite{Albuquerque:2003ei,Kavanagh:2017cru}.}. There are a number of
complementary constraints on SIMP DM, including searches for DM annihilation to neutrinos \cite{Albuquerque:2010bt}; anomalous heating of the Earth  \cite{Mack:2007xj,Mack:2012ju}; heating of Galactic gas clouds \cite{Bhoonah:2018wmw,Bhoonah:2018gjb}; and DM-cosmic ray interactions \cite{Cyburt:2002uw}. These typically require additional assumptions about the properties of the DM particle, while the current constraints depend only on its large scattering cross section with nuclei and its interactions near the Earth.

The dark matter search has also been extended to interactions via the Migdal effect, resulting in the exclusion for the first time of particles with masses between 45 and 150~MeV/c$^{2}$ with cross sections ranging from $10^{-29}$ to $10^{-26}$~cm$^2$.
These limits also take fully into account the modeling of Earth-shielding effects essential for obtaining accurate constraints for such large cross section values.

Finally, interpreted in terms of spin-dependent interactions with protons or neutrons, these results exclude new regions of the parameter space below masses of 1.3 GeV/c$^2$. In this case, atmospheric stopping is significant and has a strong effect on the derived exclusion limits, in particular for WIMP-proton interactions and the Migdal effect.

The next steps for the EDELWEISS collaboration are twofold. Such high-performance detectors are currently running in the low-background environment of its underground facility~\cite{edwtech} at the Laboratoire Souterrain de Modane  to better understand the origin of the observed events at low energy. Also, the detectors are equipped with electrodes in order to amplify the signal using the Neganov-Trofimov-Luke amplification~\cite{Neganov,Luke}, a particularly enticing prospect for searches using the electron recoils produced by the Migdal effect. The level of performance achieved in this work also opens the possibility of a first experimental measurement of the Migdal effect using a neutron calibration source.

Eventually, this level of detector performance, achieved in an above-ground laboratory with a 30~g-scale massive bolometer, is also very promising in the context of a low-energy and high precision measurement of the Coherent Elastic Neutrino-Nucleus Scattering process~\cite{Billard:2016giu}. 

\begin{acknowledgments}

We thank Felix Kahlhoefer for helpful discussions about the Migdal effect in semi-conductor materials.

BJK acknowledges funding from the Netherlands Organization for Scientific Research (NWO) through the VIDI research program ``Probing the Genesis of dark matter" (680-47-532). Part of this work was carried out on the Dutch national e-infrastructure with the support of SURF Cooperative.

The EDELWEISS project is supported in part by the German Helmholtz Alliance for Astroparticle Physics (HAP), by the French Agence Nationale pour la Recherche (ANR) and the LabEx Lyon Institute of Origins (ANR-10-LABX-0066) of the Universit\'e de Lyon within the program ``Investissements d'Avenir'' (ANR-11-IDEX-00007), by the P2IO LabEx (ANR-10-LABX-0038) in the framework ``Investissements d'Avenir'' (ANR-11-IDEX-0003-01) managed by the ANR (France), and the Russian Foundation for Basic Research (grant No. 18-02-00159).
\end{acknowledgments}

\appendix

\section{Analytic Detector Response}
\label{sec:response}
 
This appendix describes the analytic model for the detector response which is used to set limits in the SIMP and Migdal effect analyses. 
 
As described in Sec.~\ref{sec:processing}, the optimal filter (OF) algorithm is used to search for candidate events in the data stream and recover the pulse amplitude. The OF algorithm locks on the largest fluctuation away from zero in a given search window. Note that this could be a downward fluctuation, such that the reconstructed pulse amplitude may be negative. The sampling frequency is $400\,\mathrm{Hz}$ while the OF search window has a width of $20\,\mathrm{ms}$.  If each sample were independent, this would correspond to a total of 8 samples. However, correlations in the noise mean that the effective number of independent samples $N_s$ in each search window is smaller. 

We now calculate the probability distribution for the largest fluctuation (i.e.\ deviation from zero) among $N_s$ independent random variables $X_i$. We denote the probability distribution of each random variable $P_i(X_i)$ and the largest fluctuation $X^*$. With this, we obtain:
\begin{align}
\label{eq:PDFfluctuation}
P(X^*) &= \sum_i^{N_s} P_i(X^*)\prod_{k \neq i} P_k\left(|X_k| \leq X^*\right)\,.
\end{align}
This is the probability that one of the random variables lies in the interval $X^* \rightarrow X^* + \mathrm{d}X^*$, while the remaining random variables $X_{k \neq i}$ have absolute values smaller than $X^*$.

We assume Gaussian noise and fix the resolution to $\sigma = 18\,\mathrm{eV}$ (see right panel of Fig.~\ref{fig:Calibration}). 
If there are no signal events in the OF search window, then the distribution of the largest fluctuation is given by:
\begin{align}
\label{eq:Pnoise}
P(X^*) = \frac{N_s}{\sqrt{2\pi\sigma^2}} \exp \left(-\frac{|X^*|^2}{2\sigma^2}\right) \left[ \operatorname { erf } \left( \frac { |X^*| } { \sqrt { 2 }\sigma }\right) \right]^{(N_s-1)}\,.
\end{align}
Identifying $X^*$ with $E_\mathrm{out}$, we can compare this distribution with the reconstructed energy of noise-only events, for which we find that $N_s = 3$ provides a very good fit.

\begin{figure*}[tbh!]
\includegraphics[width=.99\textwidth]{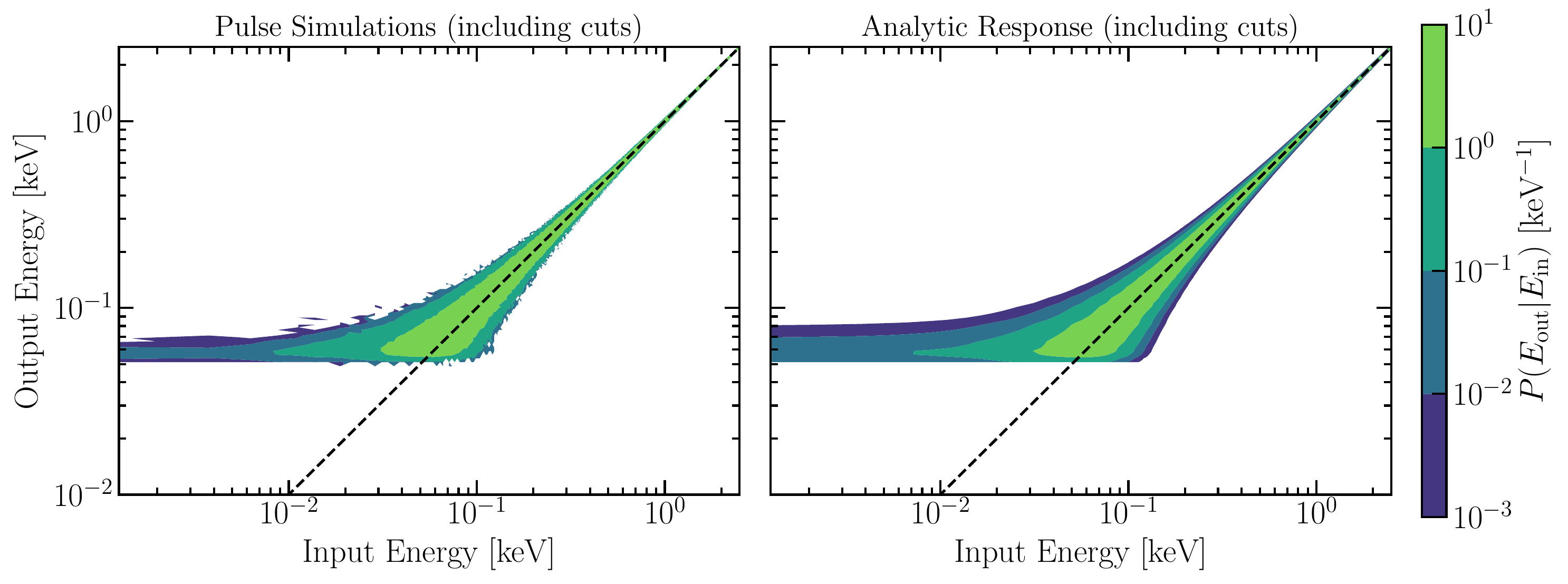}
  \caption{Comparison of pulse simulations and analytic detector response. \textbf{Left:} density function for the output energy $E_\mathrm{out}$ at a given input energy $E_\mathrm{in}$ estimated from pulse simulations after applying cuts. \textbf{Right:} analytic density function described in Eq.~\eqref{eq:Psignal}, including optimal filter resolution and efficiency cuts. We use bins of 2.5 eV in both $E_\mathrm{in}$ and $E_\mathrm{out}$.}
  \label{fig:Comparison}
\end{figure*}

If a signal event with energy $E_\mathrm{in}$ appears in one of these $N_s = 3$ independent samples, the resulting random variable will still be Gaussian-distributed with width $\sigma$ but now centred on $E_\mathrm{in}$ (the remaining two random variables are Gaussian-distributed with mean zero). Again, applying Eq.~\ref{eq:PDFfluctuation}, we obtain the probability distribution for the largest fluctuation:
\begin{widetext}
\begin{align}
\begin{split}
\label{eq:Psignal}
P_\mathrm{OF}(X^*|E_\mathrm{in}) &= \frac{1}{\sqrt{2\pi\sigma^2}} \exp \left(-\frac{(X^* - E_\mathrm{in})^2}{2\sigma^2}\right) \left[ \operatorname { erf } \left( \frac { |X^*| } { \sqrt { 2 }\sigma }\right) \right]^{(N_s-1)}\\
& +\frac{(N_s-1)}{\sqrt{2\pi\sigma^2}} \exp \left(-\frac{|X^*|^2}{2\sigma^2}\right) \left[ \operatorname { erf } \left( \frac { |X^*| } { \sqrt { 2 }\sigma }\right) \right]^{(N_s-2)} \frac{1}{2}\left[ \operatorname { erf } \left( \frac { |X^*| - E_\mathrm{in} } { \sqrt { 2 }\sigma }\right) - \operatorname { erf } \left( \frac { -|X^*| - E_\mathrm{in} } { \sqrt { 2 }\sigma }\right)  \right]\,.
\end{split}
\end{align}
\end{widetext}
At small input energies, this expression accounts for the fact that the largest fluctuation in the OF search window may not correspond to the signal pulse. At large input energies, we recover a standard Gaussian distribution for the output energy.

Finally, we apply the trigger efficiency $\eta(E_\mathrm{out})$ as a function of the reconstructed energy, given by the red curve in the right panel of Fig.~\ref{fig:Trigger}. The final detector response is then:
\begin{align}
P(E_\mathrm{out}|E_\mathrm{in}) = \eta(E_\mathrm{out}) P_\mathrm{OF}(E_\mathrm{out}|E_\mathrm{in})\,.
\end{align}
The observed spectrum of events is then obtained by a convolution:
\begin{align}
    \frac{\mathrm{d}R}{\mathrm{d}E_\mathrm{out}} = \eta(E_\mathrm{out})\int_0^\infty  P_\mathrm{OF}(E_\mathrm{out}|E_\mathrm{in}) \frac{\mathrm{d}R} {\mathrm{d}E_\mathrm{in}}\,\mathrm{d}E_\mathrm{in}\,.
\end{align}

In Fig.~\ref{fig:Comparison}, we compare the probability distribution for the output energy (at fixed input energy) estimated using pulse simulations (left panel) and using the analytic prescription (right panel). We use more than $10^7$ pulse simulations, which were performed in order to calculate the standard WIMP limits presented in Sec.~\ref{sec:WIMP}. We include only those simulated events passing all cuts and present the results in bins of 2.5 eV in both $E_\mathrm{in}$ and $E_\mathrm{out}$. The analytic response function captures the main features of detector response seen in the pulse simulations.

We have also performed a more quantitative comparison. Binning the events in bins of 1 eV in input energy, we performed a standard Kolmogorov-Smirnov (KS) test \cite{Kolmogorov,Smirnov} in each bin. We compared the output energy distribution of pulse simulations with the distribution expected from the analytic model. We found that the two distributions are consistent, with $p$-values ranging from $10^{-3}$ to 1 over the 2000 bins as expected.

Finally, we have verified that the standard WIMP limits calculated using this analytic model are in good agreement with those obtained using the pulse simulations (as described in Sec.~\ref{sec:WIMP}). We find that the two methods agree at the 10\% level.

\newcommand{\arXiv}[1]{\href{https://arxiv.org/abs/#1}{arXiv:#1}}
\newcommand{\oldarXiv}[1]{\href{https://arxiv.org/abs/#1}{#1}}
\newcommand{\DOI}{https://doi.org}

\frenchspacing


\begin{thebibliography}{100}

\bibitem{planck2018-vi} 
N. Aghanim {\it et al.}  (Planck Collaboration),
 submitted  A\&A (2018), \arXiv{1807.06209}.

\bibitem{Jungman} 
G. Jungman, M. Kamionkowski and K. Griest, \href{\DOI/10.1016/0370-1573(95)00058-5}{Phys. Rep. {\bf 267}, 195 (1996)}, \oldarXiv{hep-ph/9506380}.

\bibitem{Bertone} 
G.~Bertone and D.~Hooper,
  \href{\DOI/10.1103/RevModPhys.90.045002}{Rev.\ Mod.\ Phys.\  {\bf 90}, no. 4, 045002 (2018)}, \arXiv{1605.04909}.

\bibitem{Goodman:1984dc} 
M.~W.~Goodman and E.~Witten,
  \href{\DOI/10.1103/PhysRevD.31.3059}{Phys.\ Rev.\ D {\bf 31}, 3059 (1985)}.

\bibitem{annual} 
A. K. Drukier, K. Freese and D. N. Spergel, \href{\DOI/10.1103/PhysRevD.33.3495}{Phys. Rev. D {\bf 33}, 3495 (1986)}.

\bibitem{xenon1t} 
E. Aprile {\it et al.} (XENON Collaboration),
\href{\DOI/10.1103/PhysRevLett.121.111302}{Phys.\ Rev.\ Lett.\ {\bf 121}, 111302 (2018)}, \arXiv{1805.12562}.

\bibitem{lux} 
D. S. Akerib {\it et al.} (LUX Collaboration),
\href{\DOI/10.1103/PhysRevLett.118.021303}{Phys.\ Rev.\ Lett.\ {\bf 118}, 021303 (2017)}, \arXiv{1608.07648}.

\bibitem{pandax} 
A. Tan {\it et al.} (PandaX-II Collaboration),
\href{\DOI/10.1103/PhysRevLett.117.121303}{Phys.\ Rev.\ Lett.\ \textbf{117}, 121303 (2016)}, \arXiv{1607.07400}.

\bibitem{Essig} 
R. Essig, J. Kaplan, P. Schuster and N. Toro, (2010),
\arXiv{1004.0691}.

\bibitem{Cheung} 
C. Cheung, J. T. Ruderman, L.-T. Wang and I. Yavin, 
\href{\DOI/10.1103/PhysRevD.80.035008}{Phys.\ Rev.\ D {\bf 80}, 035008 (2009)}, \arXiv{0902.3246}.

\bibitem{Hooper} 
D. Hooper and W. Xue, 
\href{\DOI/10.1103/PhysRevLett.110.041302}{Phys. Rev. Lett. {\bf 110}, 041302 (2013)}, \arXiv{1210.1220}.

\bibitem{Falkowski} 
A. Falkowski, J.T. Ruderman and T. Volansky,
\href{\DOI/10.1007/JHEP05(2011)106}{JHEP {\bf 1105}, 106 (2011)}, \arXiv{1101.4936}.

\bibitem{Petraki} 
K. Petraki and R.R. Volkas, 
\href{\DOI/10.1142/S0217751X13300287}{Int. J. Mod. Phys. A {\bf 28}, 1330028 (2013)}, \arXiv{1305.4939}.

\bibitem{Zurek} 
K. M. Zurek, 
\href{\DOI/10.1016/j.physrep.2013.12.001}{Phys. Rep. {\bf 537}, 91 (2014)}, \arXiv{1308.0338}.

\bibitem{Bertone:2018krk} 
G.~Bertone and T.~M.~P.~Tait,
  \href{\DOI/10.1038/s41586-018-0542-z}{Nature {\bf 562}, 51-56 (2018)}, \arXiv{1810.01668}.

\bibitem{cresst} 
G. Angloher {\it et al.} (CRESST Collaboration), \href{\DOI/10.1140/epjc/s10052-016-3877-3}{Eur. Phys. J. C {\bf 76}, 25 (2016)}, \arXiv{1509.01515}.

\bibitem{cdmslite} 
R. Agnese {\it et al.} (SuperCDMS Collaboration),
\href{\DOI/10.1103/PhysRevD.97.022002}{Phys.\ Rev.\ D {\bf 97}, 022002 (2018)}, \arXiv{1707.01632}.

\bibitem{NTD} 
E.E. Haller, K.M. Itoh, J.W. Beeman, W.L. Hansen and V.I. Ozhogin, \href{\DOI/10.1117/12.176771}{SPIE 2198, Instrumentation in Astronomy VIII (1994) 630}.

\bibitem{Starkman:1990nj} 
G.~D.~Starkman, A.~Gould, R.~Esmailzadeh and S.~Dimopoulos,
  \href{\DOI/10.1103/PhysRevD.41.3594}{Phys.\ Rev.\ D {\bf 41}, 3594 (1990)}.

\bibitem{Collar1992} 
J.~I.~Collar and F.~T.~Avignone III, \href{https://doi.org/10.1016/0370-2693(92)90873-3}{Phys.\ Lett.\ B \textbf{275}, 181 (1992)}.

\bibitem{Collar1993} 
J.~I.~Collar and F.~T.~Avignone III, \href{https://doi.org/10.1103/PhysRevD.47.5238}{Phys.\ Rev.\ D \textbf{47}, 5238 (1993)}.

\bibitem{Hasenbalg:1997hs} 
F.~Hasenbalg {\it et al.},
  \href{\DOI/10.1103/PhysRevD.55.7350}{Phys.\ Rev.\ D {\bf 55}, 7350 (1997)}, \oldarXiv{astro-ph/9702165}.

\bibitem{Kouvaris:2014lpa} 
C.~Kouvaris and I.~M.~Shoemaker,
  \href{\DOI/10.1103/PhysRevD.90.095011}{Phys.\ Rev.\ D {\bf 90}, 095011 (2014)}, \arXiv{1405.1729}.

\bibitem{Kouvaris:2015laa} 
C.~Kouvaris,
  \href{\DOI/10.1103/PhysRevD.93.035023}{Phys.\ Rev.\ D {\bf 93}, 035023 (2016)}, \arXiv{1509.08720}.

\bibitem{Bernabei:2015nia} 
R.~Bernabei {\it et al.},
  \href{\DOI/10.1140/epjc/s10052-015-3473-y}{Eur.\ Phys.\ J.\ C {\bf 75}, 239 (2015)}, \arXiv{1505.05336}.

\bibitem{Kavanagh:2016pyr} 
B.~J.~Kavanagh, R.~Catena and C.~Kouvaris,
  \href{\DOI/10.1088/1475-7516/2017/01/012}{JCAP {\bf 1701} (2017) 012}, \arXiv{1611.05453}.

\bibitem{Hochberg:2014dra} 
Y.~Hochberg, E.~Kuflik, T.~Volansky and J.~G.~Wacker,
  \href{\DOI/10.1103/PhysRevLett.113.171301}{Phys.\ Rev.\ Lett.\  {\bf 113}, 171301 (2014)}, \arXiv{1402.5143}.

\bibitem{Hochberg:2014kqa} 
Y.~Hochberg, E.~Kuflik, H.~Murayama, T.~Volansky and J.~G.~Wacker,
  \href{\DOI/10.1103/PhysRevLett.115.021301}{Phys.\ Rev.\ Lett.\  {\bf 115}, 021301 (2015)}, \arXiv{1411.3727}.

\bibitem{Albuquerque:2003ei} 
I.~F.~M.~Albuquerque and L.~Baudis,
  \href{\DOI/10.1103/PhysRevLett.90.221301}{Phys.\ Rev.\ Lett.\  {\bf 90}, 221301 (2003)}, Erratum: [\href{\DOI/10.1103/PhysRevLett.91.229903}{Phys.\ Rev.\ Lett.\  {\bf 91}, 229903 (2003)}], \oldarXiv{astro-ph/0301188}.

\bibitem{Davis:2017noy} 
J.~H.~Davis,
  \href{\DOI/10.1103/PhysRevLett.119.211302}{Phys.\ Rev.\ Lett.\  {\bf 119}, 211302 (2017)}, \arXiv{1708.01484}.

\bibitem{Kavanagh:2017cru} 
B.~J.~Kavanagh,
  \href{\DOI/10.1103/PhysRevD.97.123013}{Phys.\ Rev.\ D {\bf 97}, 123013 (2018)}, \arXiv{1712.04901}.

\bibitem{Vergados:2004bm} 
J.~D.~Vergados and H.~Ejiri,
  \href{\DOI/10.1016/j.physletb.2004.11.085}{Phys.\ Lett.\ B {\bf 606}, 313 (2005)}, \oldarXiv{hep-ph/0401151}.

\bibitem{Moustakidis:2005gx} 
C.~C.~Moustakidis, J.~D.~Vergados and H.~Ejiri,
  \href{\DOI/10.1016/j.nuclphysb.2005.08.033}{Nucl.\ Phys.\ B {\bf 727}, 406 (2005)}, \oldarXiv{hep-ph/0507123}.

\bibitem{Bernabei:2007jz} 
R.~Bernabei {\it et al.},
  \href{\DOI/10.1142/S0217751X07037093}{Int.\ J.\ Mod.\ Phys.\ A {\bf 22}, 3155 (2007)}, \arXiv{0706.1421}.

\bibitem{Ibe:2017yqa} 
M.~Ibe, W.~Nakano, Y.~Shoji and K.~Suzuki,
  \href{\DOI/10.1007/JHEP03(2018)194}{JHEP {\bf 1803}, 194 (2018)}, \arXiv{1707.07258}.

\bibitem{Dolan:2017xbu} 
M.~J.~Dolan, F.~Kahlhoefer and C.~McCabe,
  \href{\DOI/10.1103/PhysRevLett.121.101801}{Phys.\ Rev.\ Lett.\  {\bf 121}, 101801 (2018)}, \arXiv{1711.09906}.

\bibitem{Akerib:2018hck} 
D.~S.~Akerib {\it et al.} (LUX Collaboration), \arXiv{1811.11241}.

\bibitem{pyle} 
M.~Pyle, E.~Feliciano-Figueroa and B.~Sadoulet,
  \arXiv{1503.01200}.

\bibitem{edwtech} 
E. Armengaud {\it et al.} (EDELWEISS Collaboration), 
\href{\DOI/10.1088/1748-0221/12/08/P08010}{JINST {\bf 12}, P08010 (2017)}, \arXiv{1706.01070}.

\bibitem{quenching-edw} 
A. Benoit {\it et al.} (EDELWEISS Collaboration), \href{\DOI/10.1016/j.nima.2007.04.118}{Nucl. Instrum. Meth. A {\bf 577}, 558 (2007)}, \oldarXiv{astro-ph/0607502}.

\bibitem{quenching-cdms} 
R. Agnese {\it et al.} (SuperCDMS Collaboration), \href{\DOI/10.1063/1.5041457}{Appl. Phys. Lett. {\bf 113}, 092101 (2018)}, \arXiv{1805.09942}.

\bibitem{cryoconcept} 
http://cryoconcept.com/, (accessed on March 28th, 2019).

\bibitem{cryovib} 
E. Olivieri, J. Billard, M. De Jesus, A. Juillard, A. Leder, 
\href{\DOI/10.1016/j.nima.2017.03.045}{Nucl. Instrum. Meth. A {\bf 858}, 73 (2017)}, \arXiv{1703.08957}.

\bibitem{suspendedtower} 
R. Maisonobe {\it et al.}, 
\href{\DOI/10.1088/1748-0221/13/08/T08009}{JINST {\bf 13} (2018) T08009}, \arXiv{1803.03463}.

%

\bibitem{thermModel} 
{\it Paper in preparation}.

\bibitem{trigger} 
S.~Di Domizio, F.~Orio and M.~Vignati,
  \href{\DOI/10.1088/1748-0221/6/02/P02007}{JINST {\bf 6}, P02007 (2011)}, \arXiv{1012.1263}.

\bibitem{standardassumptions} 
J.~D.~Lewin and P.~F.~Smith, 
\href{\DOI/10.1016/S0927-6505(96)00047-3}{Astropart.\ Phys.\ {\bf 6}, 87 (1996)};
C.~Savage, K.~Freese and P.~Gondolo, 
\href{\DOI/10.1103/PhysRevD.74.043531}{Phys.\ Rev.\ D {\bf 74}, 043531 (2006)}, \oldarXiv{astro-ph/0607121}.


\bibitem{edwiii} 
L.~Hehn {\it et al.} (EDELWEISS Collaboration), 
\href{\DOI/10.1140/epjc/s10052-016-4388-y}{Eur. Phys. J. C {\bf 76} 548 (2016)}, \arXiv{1607.03367}.

\bibitem{darkside} 
P. Agnese {\it et al.} (DarkSide Collaboration), 
\href{\DOI/10.1103/PhysRevLett.121.081307}{Phys. Rev. Lett. {\bf 121}, 081307 (2018)}, \arXiv{1802.06994}.

\bibitem{nucleus} 
G.~Angloher {\it et al.} (CRESST Collaboration),
  \href{\DOI/10.1140/epjc/s10052-017-5223-9}{Eur.\ Phys.\ J.\ C {\bf 77} (2017) 637}, \arXiv{1707.06749}.

\bibitem{newsg} 
Q. Arnaud {\it et al.} (NEWS-G Coll,)
\href{\DOI/10.1016/j.astropartphys.2017.10.009}{Astropart. Phys. {\bf 97}, 54 (2018)}, \arXiv{1706.04934}.

\bibitem{xqc} 
A. L. Erickcek, P. J. Steinhardt, D. McCammon and P. C. McGuire,
\href{\DOI/10.1103/PhysRevD.76.042007}{Phys. Rev. D {\bf 76}, 042007 (2007)}, \arXiv{0704.0794}.

\bibitem{neutrinoBillard} 
J.~Billard, L.~Strigari and E.~Figueroa-Feliciano, \href{\DOI/10.1103/PhysRevD.89.023524}{Phys.\ Rev.\ D {\bf 89}, 023524 (2014)}, \arXiv{1307.5458}.

\bibitem{Mahdawi:2018euy} 
M.~S.~Mahdawi and G.~R.~Farrar,
  \href{\DOI/10.1088/1475-7516/2018/10/007}{JCAP {\bf 1810} (2018) 007}, \arXiv{1804.03073}.

\bibitem{Gluscevic:2017ywp} 
V.~Gluscevic and K.~K.~Boddy,
  \href{\DOI/10.1103/PhysRevLett.121.081301}{Phys.\ Rev.\ Lett.\  {\bf 121}, 081301 (2018)}, \arXiv{1712.07133}.

\bibitem{Hooper:2018bfw} 
D.~Hooper and S.~D.~McDermott,
  \href{\DOI/10.1103/PhysRevD.97.115006}{Phys.\ Rev.\ D {\bf 97}, 115006 (2018)}, \arXiv{1802.03025}.

\bibitem{verne} 
B.~J.~Kavanagh, ``verne v1.0 [computer software]", Astrophysics  Source  Code  Library,  record \href{http://ascl.net/1802.005}{ascl:1802.005} (archived on Zenodo, \href{\DOI/10.5281/zenodo.1115601}{DOI:10.5281/zenodo.1115601}) (2017).

\bibitem{Emken:2018run} 
T.~Emken and C.~Kouvaris,
  \href{\DOI/10.1103/PhysRevD.97.115047}{Phys.\ Rev.\ D {\bf 97}, 115047 (2018)}, \arXiv{1802.04764}.

\bibitem{Emken:2017qmp} 
T.~Emken and C.~Kouvaris,
  \href{\DOI/10.1088/1475-7516/2017/10/031}{JCAP {\bf 1710} (2017) 031}, \arXiv{1706.02249}.
  
  \bibitem{Collar:2018ydf} 
  J.~I.~Collar,
  Phys.\ Rev.\ D {\bf 98}, no. 2, 023005 (2018).

\bibitem{Mahdawi:2017cxz} 
M.~S.~Mahdawi and G.~R.~Farrar,
  \href{\DOI/10.1088/1475-7516/2017/12/004}{JCAP {\bf 1712} (2017) 004}, \arXiv{1709.00430}.

\bibitem{Mahdawi:2017utm} 
M.~S.~Mahdawi and G.~R.~Farrar, (2017), \arXiv{1712.01170}.

\bibitem{Mayet:2016zxu} 
F.~Mayet {\it et al.},
  \href{\DOI/10.1016/j.physrep.2016.02.007}{Phys.\ Rep.\  {\bf 627}, 1 (2016)}, \arXiv{1602.03781}.

\bibitem{Gould:1987ww} 
A.~Gould,
  \href{\DOI/10.1086/166347}{Astrophys.\ J.\  {\bf 328}, 919 (1988)}.

\bibitem{Mack:2007xj} 
G.~D.~Mack, J.~F.~Beacom and G.~Bertone,
  \href{\DOI/10.1103/PhysRevD.76.043523}{Phys.\ Rev.\ D {\bf 76}, 043523 (2007)}, \arXiv{0705.4298}.

\bibitem{Kouvaris:2015xga} 
C.~Kouvaris and N.~G.~Nielsen,
  \href{\DOI/10.1103/PhysRevD.92.075016}{Phys.\ Rev.\ D {\bf 92}, 075016 (2015)}, \arXiv{1505.02615}.

\bibitem{Klos:2013rwa}
  P.~Klos, J.~Men\'endez, D.~Gazit and A.~Schwenk,
  \href{\DOI/10.1103/PhysRevD.88.083516}{Phys.\ Rev.\ D {\bf 88}, 083516 (2013)}, \arXiv{1304.7684}, 
   Erratum: [\href{\DOI/10.1103/PhysRevD.89.029901}{Phys.\ Rev.\ D {\bf 89}, 029901 (2014)}].
  
\bibitem{Akerib:2017kat} 
  D.~S.~Akerib {\it et al.} (LUX Collaboration),
  Phys.\ Rev.\ Lett.\  {\bf 118}, no. 25, 251302 (2017),
  \arXiv{1705.03380}.
  
  \bibitem{Aprile:2019dbj} 
  E.~Aprile {\it et al.} (XENON Collaboration),
  \arXiv{1902.03234}.
  
\bibitem{Amole:2019fdf} 
  C.~Amole {\it et al.} (PICO Collaboration),
  \arXiv{1902.04031}.
  
  \bibitem{Agnese:2017jvy} 
  R.~Agnese {\it et al.} (SuperCDMS Collaboration),
  Phys.\ Rev.\ D {\bf 97}, no. 2, 022002 (2018).
  
  \bibitem{Fu:2016ega} 
  C.~Fu {\it et al.} (PandaX-II Collaboration),
  Phys.\ Rev.\ Lett.\  {\bf 118}, no. 7, 071301 (2017)
  Erratum: [Phys.\ Rev.\ Lett.\  {\bf 120}, no. 4, 049902 (2018)].
  
  \bibitem{Abdelhameed:2019szb} 
  A.~H.~Abdelhameed {\it et al.} (CRESST Collaboration),
  \arXiv{1902.07587}.


\bibitem{Albuquerque:2010bt} 
I.~F.~M.~Albuquerque and C.~Perez de los Heros,
  \href{\DOI/10.1103/PhysRevD.81.063510}{Phys.\ Rev.\ D {\bf 81}, 063510 (2010)}, \arXiv{1001.1381}.

\bibitem{Mack:2012ju} 
G.~D.~Mack and A.~Manohar,
  \href{\DOI/10.1088/0954-3899/40/11/115202}{J.\ Phys.\ G {\bf 40}, 115202 (2013)}, \arXiv{1211.1951}.

\bibitem{Bhoonah:2018wmw} 
A.~Bhoonah, J.~Bramante, F.~Elahi and S.~Schon,
  \href{\DOI/10.1103/PhysRevLett.121.131101}{Phys.\ Rev.\ Lett.\  {\bf 121}, 131101 (2018)}, \arXiv{1806.06857}.

\bibitem{Bhoonah:2018gjb} 
A.~Bhoonah, J.~Bramante, F.~Elahi and S.~Schon,
  \arXiv{1812.10919}.

\bibitem{Cyburt:2002uw} 
R.~H.~Cyburt, B.~D.~Fields, V.~Pavlidou and B.~D.~Wandelt,
  \href{\DOI/10.1103/PhysRevD.65.123503}{Phys.\ Rev.\ D {\bf 65}, 123503 (2002)}, \oldarXiv{astro-ph/0203240}.

\bibitem{Neganov} 
B. Neganov and V. Trofimov, Otkryt.\ Izobret.\ {\bf 146}, 215 (1985).

\bibitem{Luke} 
P. N. Luke, J.\ Appl.\ Phys.\ {\bf 64}, 6858 (1988).

\bibitem{Kolmogorov} 
A.~Kolmogorov, G.\ Ist.\ Ital.\ Attuari.\ \textbf{4}:1, 83 (1933).

\bibitem{Smirnov} 
 N.~Smirnov, \href{\DOI/10.1214/aoms/1177730256}{Ann.\ Math.\ Statist.\ \textbf{19}, 2 (1948), 279}.

\bibitem{Billard:2016giu} 
  J.~Billard {\it et al.},
  J.\ Phys.\ G {\bf 44}, no. 10, 105101 (2017).
  

  

  
  

\end{thebibliography}
\end{document}